\documentclass{article} 
\usepackage{graphicx}
\usepackage{color}
\usepackage{subfig}
\usepackage{amssymb}
\usepackage{amsthm}
\usepackage{amsmath} 
\usepackage{mathrsfs}

\begin{document}
\newcommand{\erfc}{\mathrm{erfc}}

\title{Classical and Quantum Chaos in the Diamond Shaped Billiard}
\author{R. Salazar, G.~T\'ellez, D. Jaramillo \\
Departamento de F\'{\i}sica, Universidad de los Andes\\
A.A. 4976, Bogot\'a Colombia \\D. L. Gonz\'alez \\ Department of Physics, University of Maryland \\
  College Park, Maryland 20742-4111 USA 
}

\date{}


\maketitle

\begin{abstract}
We analyse the classical and quantum behaviour of a particle trapped
in a diamond shaped billiard. We defined this billiard as 
a half stadium connected with a triangular billiard. A parameter $\xi$ which
gradually change the shape of the billiard from a regular
equilateral triangle ($\xi=1$) to a diamond 
($\xi=0$) was used to control the transition between the regular and
chaotic regimes. The classical behaviour is regular when the control
parameter $\xi$ is one; in contrast, the system is chaotic when $\xi
\neq 1$ even for values of $\xi$ close to one. The entropy grows
fast as $\xi$ is decreased from 1 and the Lyapunov exponent remains
positive for $\xi<1$. The \textit{Finite Difference Method} was
implemented in order to solve the quantum problem. The energy spectrum
and eigenstates were numerically computed for different values of the
control parameter. The nearest-neighbour spacing distribution is
analysed as a function of $\xi$, finding a \textit{Poisson} and a
\textit{Gaussian Orthogonal Ensemble} (GOE) distribution for regular
and chaotic regimes respectively. Several scars and
bouncing ball states are shown with their corresponding classical
periodic orbits. Along the document the classical chaos identifiers
are computed to show that system is chaotic. On the other hand, the
quantum counterpart is in agreement with the
\textit{Bohigas-Giannoni-Schmit} conjecture and exhibits the standard
features for chaotic billiard such as the scarring of the
wavefunction.
\end{abstract}


\section{Introduction}
A billiard is a system where a particle, with mass $m$, is trapped in
a region $\mathfrak{D}$ with perfect reflecting walls. The dynamics of
the particle varies depending on the shape of the billiard
boundary $\partial\mathfrak{D}$. The cardioid billiard, the Bunimovich
billiard (stadium billiard) and non-equilateral triangular billiards
are typical examples which exhibit classical chaos. The quantum problem is
reduced to solve the Hemholtz equation for the wave function $\psi(\vec{r})$
\begin{equation}
  \left(\vec{\nabla}^2 + \kappa^2\right)\psi(\vec{r}) = 0 \hspace{0.5cm} \mathrm{for} \hspace{0.5cm} \vec{r}  \in \mathfrak{D} 
\end{equation}
with the Dirichlet boundary condition $\psi(\vec{r}) = 0$ if $\vec{r}  \in \partial\mathfrak{D}$ where $\kappa = \sqrt{2mE}/\hbar$ is the wave vector and $E$
is the energy. The energy level statistical properties of several
Hamiltonian systems may be studied borrowing results from the random
matrices theory. For example, it is well known that the energy level spacing
distributions of a system is Poissonian if its classical counterpart
exhibits a regular motion. This is the case of billiards whose shape
is a rectangle (particle in a two dimensional box), an equilateral
triangle, a circle or an ellipse. On the other hand, if the classical
counterpart has a chaotic motion, then the energy levels follow a
\textit{Gaussian Orthogonal Ensemble} (GOE)
distribution~\cite{Mehta}. Other non convex and chaotic two
dimensional quantum cavities are the Sinai and the annular
billiards. They introduce an inner disk of infinite potential into the
rectangular and circular billiards respectively.
\begin{figure}[h]
\centering
\includegraphics[width=0.4\textwidth]{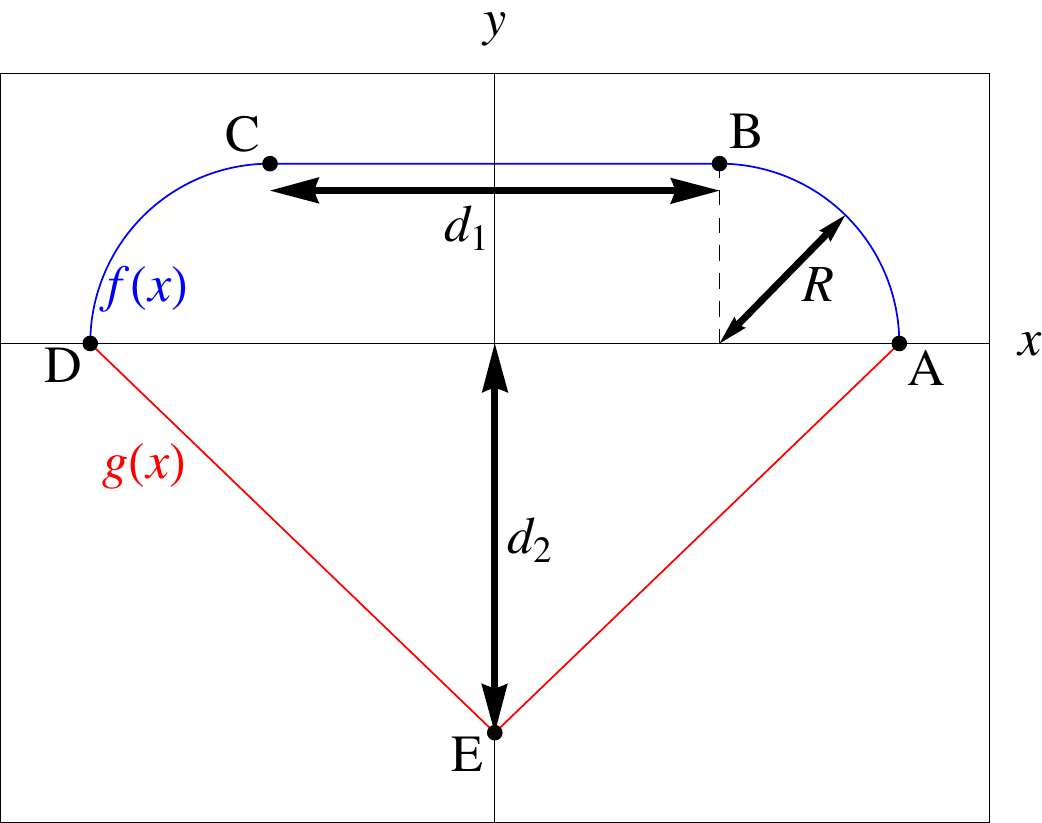}
\includegraphics[width=0.35\textwidth]{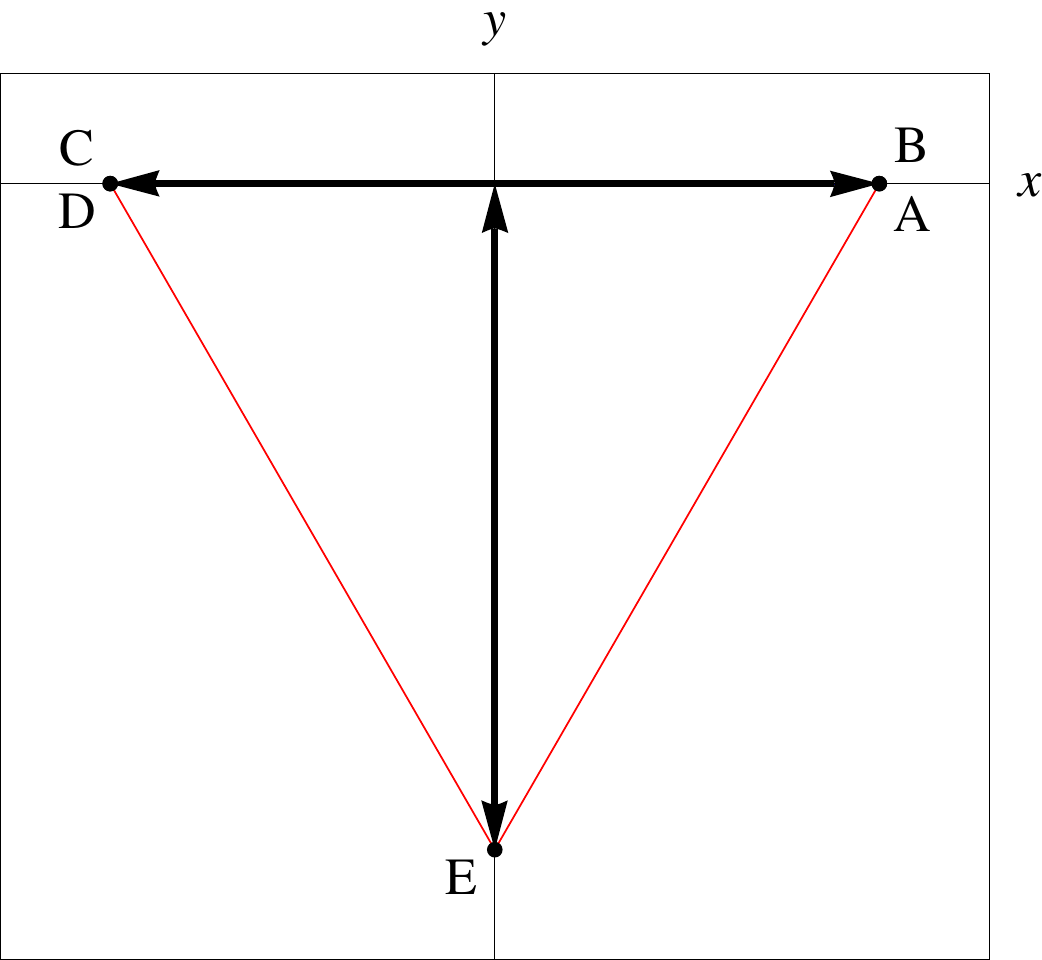}\\
\caption{\textit{(left)} The diamond shaped billiard
  is defined through the functions $f(x)$ and $g(x)$. In turn, these
  functions depend on the parameters $R$, $d_1$ and $d_2$. \textit{(right)} Triangular billiard.}   
  \label{boundaryFig}
\end{figure}\\

The system studied in this paper is the diamond shaped billiard (see Figure \ref{boundaryFig}). The upper boundary of the billiard is a half stadium defined by the equation $y=f(x)$ with
\begin{equation}
f(x) = \left\{ \begin{array}{ll}
 \sqrt{R^2-\left(x+\frac{d_1}{2}\right)^2} &\mbox{ if $-R-\frac{d_1}{2}<x\leq-\frac{d_1}{2}$} \\
 R &\mbox{ if $|x|\leq\frac{d_1}{2}$} \\
 \sqrt{R^2-\left(x-\frac{d_1}{2}\right)^2} &\mbox{ if $\frac{d_1}{2}<x\leq\frac{d_1}{2}+R$} 
       \end{array} \right. 
\end{equation}
and the lower boundary as $y=g(x)$ with
\begin{equation}
g(x) = \left\{ \begin{array}{rl}
\displaystyle
 \frac{-d_2 x}{R+\frac{d_1}{2}}-d_2 &\mbox{ if $-R-\frac{d_1}{2}<x<0$} \\
\displaystyle
 \frac{d_2 x}{R+\frac{d_1}{2}}-d_2 &\mbox{ if $0<x<R+\frac{d_1}{2}$} 
       \end{array} \right. 
\end{equation}
The parameters which define the billiard depend on the
control parameter $\xi$ according to: $R(\xi) = R_o\left(1 -
\xi\right)$, $d_1(\xi) = \left(\frac{5}{2} + \xi\right)R_o$ and
$d_2(\xi)=\sqrt{\frac{3}{4}}d_1(\xi)$. The value of $R_o$ has been
taken as one and the non-dimensional parameter $\xi$ varies in the
interval $0\leq\xi\leq1$. The shape changes from a diamond to an
equilateral triangle as the parameter $\xi$ goes from 0 to 1. The boundary may be conveniently expressed
in polar coordinates as follows
\begin{equation}
r_c(\phi) = \left\{ \begin{array}{rl}
r_+(\phi) &\mbox{ if $\phi_A\leq\phi< \phi_B$ } \\
\frac{R}{\sin\phi} &\mbox{ if $\phi_B\leq\phi< \phi_C$} \\
r_-(\phi) &\mbox{ if $\phi_C\leq\phi< \phi_D$}\\ 
l_-(\phi) &\mbox{ if $\phi_D\leq\phi< \phi_E$}\\ 
l_+(\phi) &\mbox{ if $\phi_E\leq\phi< 2\pi$} 
\end{array} \right.
\end{equation}
where the left $(-)$ and right $(+)$ quarter of circles are defined by  
\begin{equation}
r_\pm(\phi) = \frac{1}{2}\left(\pm d_1\cos\phi+\sqrt{4R^2-d_1^2\sin^2\phi}\right).
\end{equation}
The lower points $\left(x, g(x)\right)$ are located at
\begin{equation}
l_\pm(\phi) = \frac{d_2}{\frac{\mp d_2\cos\phi}{R+d_1/2}-\sin\phi}
\end{equation}
and the $\phi$-coordinate of the points from $A$ to $E$ are: $\phi_A = 0$, $\phi_B=\arctan\left(\frac{2R}{d_1}\right)$, $\phi_C=\arctan\left(\frac{2R}{d_1}\right)+2\arctan\left(\frac{d_1}{2R}\right)$, $\phi_D=\pi$ and $\phi_E=\frac{3}{2}\pi$, respectively.

\section{Classical diamond billiard}
\subsection{Trajectories}
There are two degrees of freedom in the diamond billiard, thus its
phase space has four dimensions. Because of the conservation of energy
the number of dimensions is reduced to three. Commonly, in order to
obtain the dynamical information of the system, we construct the
Poincar\'e section, and so we may study a two-dimensional map. This is
equivalent to choose two variables which define where and how the
collisions occur in the billiard. 
\begin{figure}[h]
  \centering
  \includegraphics[scale=0.4]{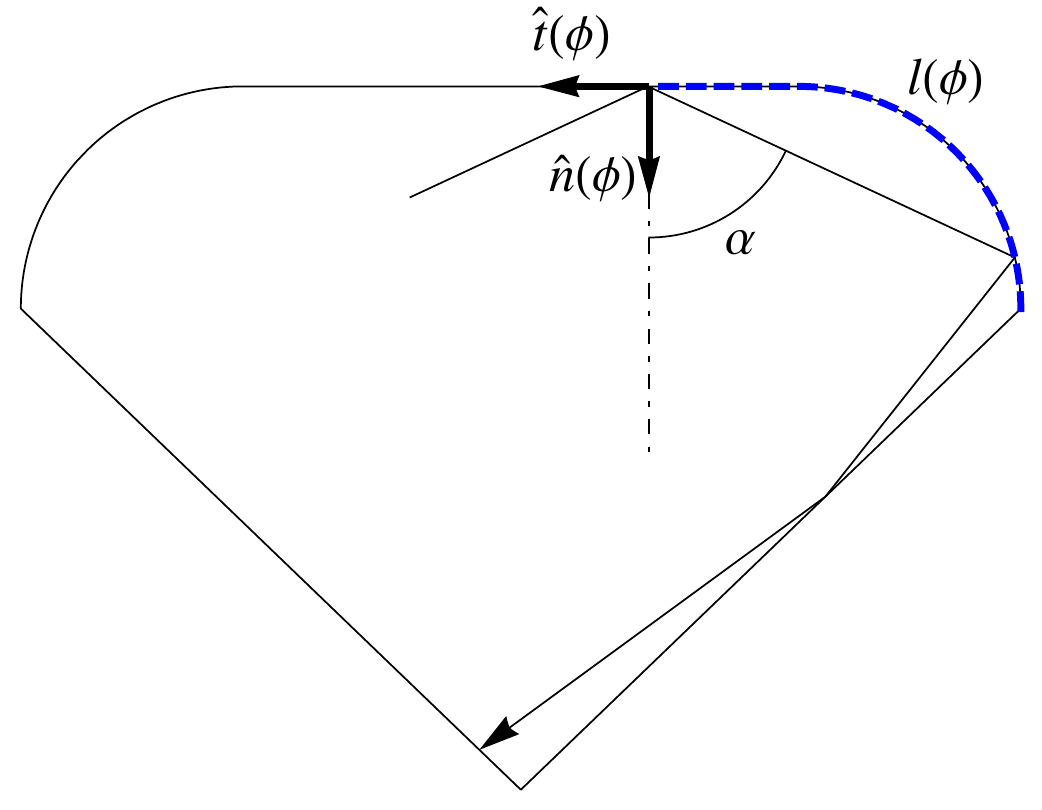} 
  \caption[Variables of the reduced phase space]%
  {Variables of the reduced phase space. The blue dashed line represents $l(\phi)$ while $\alpha$ is the angle formed by the normal vector at position $\phi$, $n(\phi)$, and the trajectory of the particle after the impact.}
  \label{phaseSpaceConvenionFig}
\end{figure}\\
We chose a rescaled arclength
$l(\phi)$, and the angle $\alpha$ which defines the direction after
the impact as the variables to describe the particle motion in the billiard. These variables are shown in Figure \ref{phaseSpaceConvenionFig}. The rescaled arclength is defined as $l(\phi)=\frac{L(\phi)}{L(2\pi)}$
where $L(\phi)=\int_0^\phi r_c(\phi) d\phi$. In order to compute the
position and velocity of the $(n+1)$-th collision, using the position
$\left(x^{(n)},
y^{(n)}\right)=\left(r_c\left(\phi^{(n)}\right),\phi^{(n)}\right)$ and
the \textit{incident} velocity $\vec{v}^{(n)}=(v_x^{(n)},v_y^{(n)})$
of the $n$-th collision, we may proceed as follows: \noindent \textit{(i) Incident velocity of the (n+1) collision.} The
normal $v_n^{(n)}$ and tangent $v_t^{(n)}$ components of the velocity
$\vec{v}^{(n)}$ are computed by projecting it into the normal,
$\hat{n}(\phi)$, and the tangent, $\hat{t}(\phi)$, unitary vectors of
the boundary. Just after the collision with the boundary the normal
component velocity changes its sign while the tangent component
remains unchanged, thus one can obtain the velocity $\vec{v}^{(n+1)}$
after the $n$-th collision which is also the incident velocity
$(n+1)$-th collision. For this calculation the components of the tangent vector
$\vec{t}=\frac{d}{d\phi}\left(r_c(\phi)\cos\phi,r_c(\phi)\sin\phi\right)=\left(T_x,T_y\right)$
are needed. These are
\begin{equation} T_x(\phi) = \left\{ \begin{array}{ll}
    -r_+(\phi)\sin\phi+\cos\phi w_+(\phi) &\mbox{ if $\phi_A<\phi<
      \phi_B$ } \\ -R &\mbox{ if $\phi_B<\phi< \phi_C$}
    \\ -r_-(\phi)\sin\phi+\cos\phi w_-(\phi) &\mbox{ if $\phi_C<\phi<
      \phi_D$}\\ \left(R+d_1/2\right)a &\mbox{ if $\phi_D<\phi< 2\pi$}
\end{array} \right.
\end{equation}
and
\begin{equation}
T_y(\phi) = \left\{ \begin{array}{ll}
r_+(\phi)\cos\phi+\sin\phi w_+(\phi) &\mbox{ if $\phi_A<\phi< \phi_B$ } \\
0 &\mbox{ if $\phi_B<\phi< \phi_C$} \\
r_-(\phi)\cos\phi+\sin\phi w_-(\phi) &\mbox{ if $\phi_C<\phi< \phi_D$}\\ 
-d_2a &\mbox{ if $\phi_D<\phi< \phi_E$}\\ 
d_2a &\mbox{ if $\phi_E<\phi< 2\pi$} 
\end{array} \right.
\end{equation}
where
\begin{equation}
w_\pm(\phi)=\mp\frac{1}{2}d_1\sin\phi\left(1 \pm \frac{d_1\cos\phi}{2r_\pm(\phi) \mp d_1\cos\phi}\right)
\end{equation}
and
\begin{equation}
a = \frac{1}{\sqrt{\left(R+\frac{d_1}{2}\right)^2 + d_2^2}}.
\end{equation}
\\The normal vector is obtained by rotating the tangent vector
$\hat{n}=\mathbb{R}(-\pi/2)\hat{t}$, hence $n_x(\phi) = t_y(\phi)$ and
$n_y(\phi) = -t_x(\phi)$.
\noindent \textit{(ii) Position of the $(n+1)$-th collision.} If the line which crosses through the points $\left(x^{(n)}, y^{(n)}\right)$ and $\left(x^{(n+1)}, y^{(n+1)}\right)$ is $Y^{(n)}(x) = m^{(n)}x+b^{(n)}$, then the slope and the $y$-intercept are
\begin{equation}
m^{(n)} =  \frac{v_y^{(n)}}{v_x^{(n)}} \hspace{0.5cm}
\mathrm{and} \hspace{0.5cm} b^{(n)} =
y^{(n)}-\frac{v_y^{(n)}}{v_x^{(n)}}y^{(n)}
\,,
\end{equation}
respectively. The intersections of a line $Y(x) = mx+b$ with the boundary are
\begin{equation}
x^*_i = \left\{ \begin{array}{rl}
x_{+}^\pm & \mbox{ if $\frac{d_1}{2}<x_{+}^\pm\leq\frac{d_1}{2}+R$ and $Y(x_{+}^\pm)>0$} \\
\frac{R-b}{m} &\mbox{ if $|\frac{R-b}{m}|\leq\frac{d_1}{2}$ and $Y(\frac{R-b}{m})>0$} \\
x_{-}^\pm &\mbox{ if $-R-\frac{d_1}{2}<x_{-}^\pm\leq-\frac{d_1}{2}$ and $Y(x_{-}^\pm)>0$} \\
x_+&\mbox{ if $-R-\frac{d_1}{2}<x_+<0$ and $Y(x_{+})<0$} \\
x_- &\mbox{ if $0<x_-<R+\frac{d_1}{2}$ and $Y(x_{-})<0$} 
\end{array} \right.
\label{eqRoots}
\end{equation}
\\where we have defined

\begin{equation}
x_\pm = \mp \frac{(b+d_2)(d_1+2R)}{2d_2 \pm m(d_1 + 2R)} \hspace{1.0cm}\mbox{and}
\end{equation}
\begin{equation}
x_{s}^{\pm} = \frac{s\hspace{0.1cm} d_1-2bm \pm \sqrt{4(1+m^2)R^2 - (d_1m + 2b)^2}}{2(1+m^2)}
\end{equation}
\\with $s=\left\{+,-\right\}$. The diamond billiard is a convex billiard, so the equation (\ref{eqRoots}) gives us two roots: $x^*_1$ and $x^*_2$, one of them is the position of the current collision so it is known, let us call it $x^{n} = (x^*_1)^{(n)}$, the other root gives the position of the $(n+1)$ collision
\begin{equation}
x^{(n+1)} = (x^*_2)^{(n)} \hspace{0.5cm} \mathrm{and} \hspace{0.5cm} y^{(n+1)} = Y^{(n)}\left((x^*_2)^{(n)}\right).
\end{equation}\\
\begin{figure}[h]
  \centering
  \includegraphics[width=0.13\textwidth]{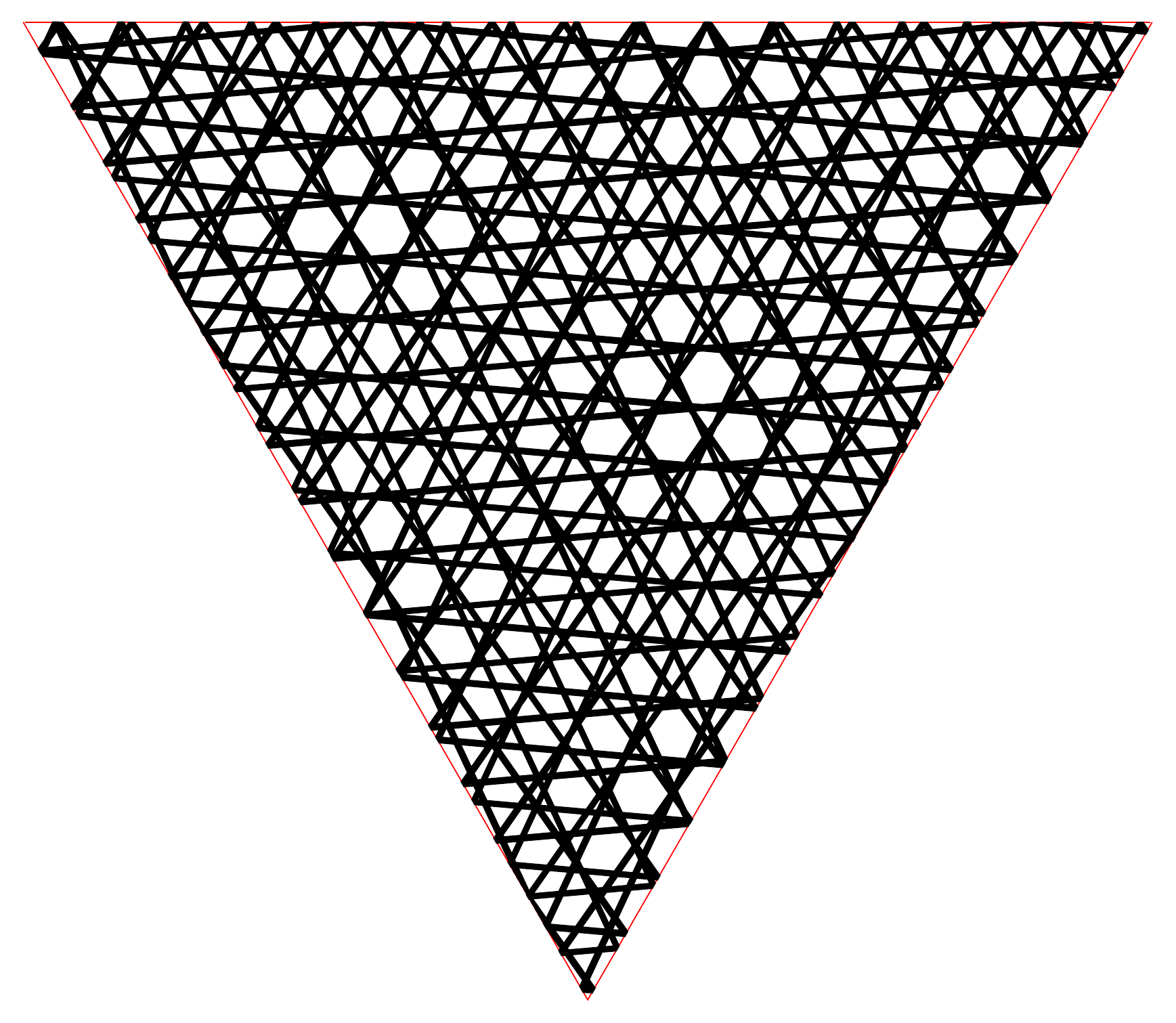}\hspace{0.5cm}
  \includegraphics[width=0.13\textwidth]{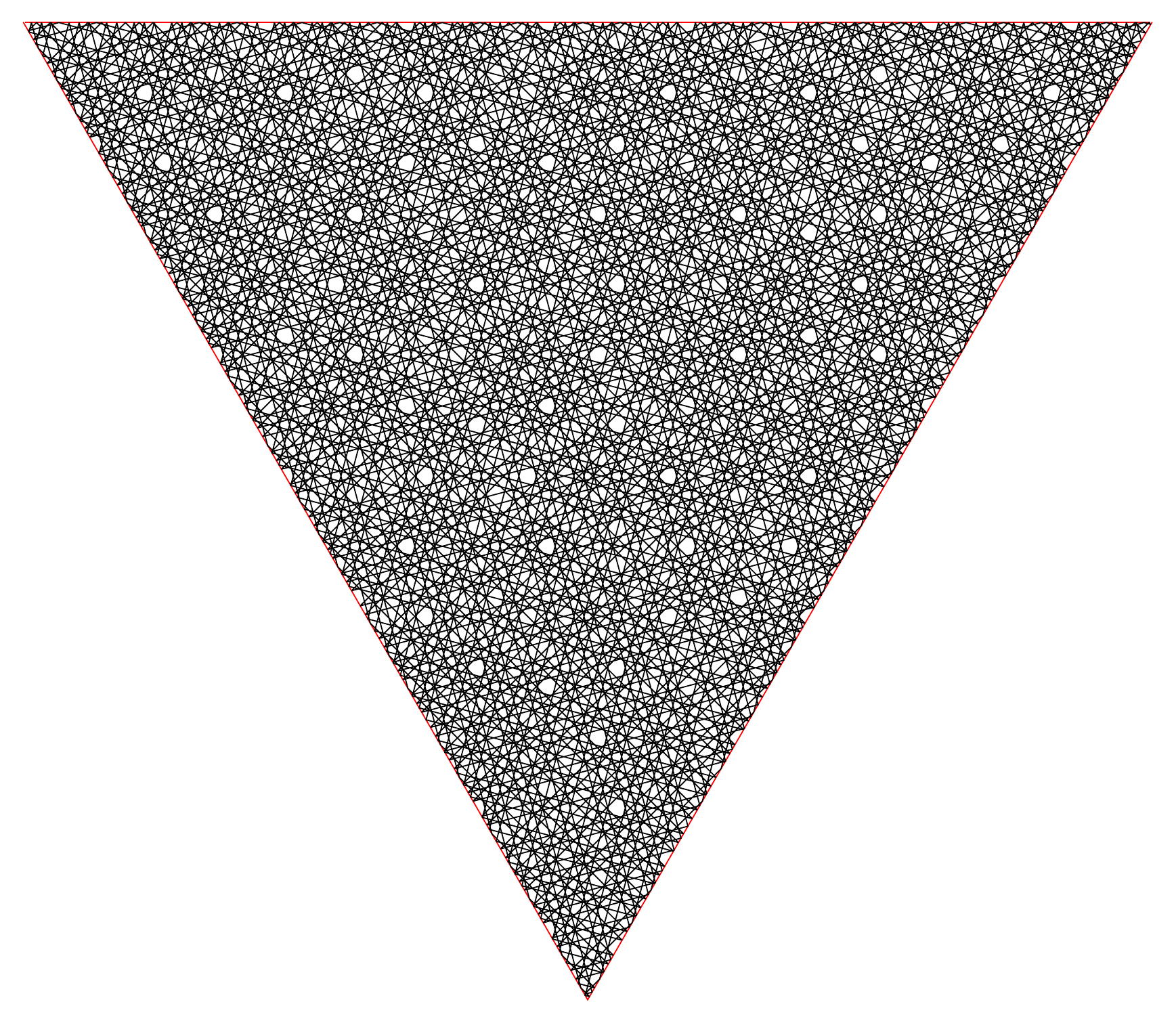}\hspace{0.5cm}
  \includegraphics[width=0.13\textwidth]{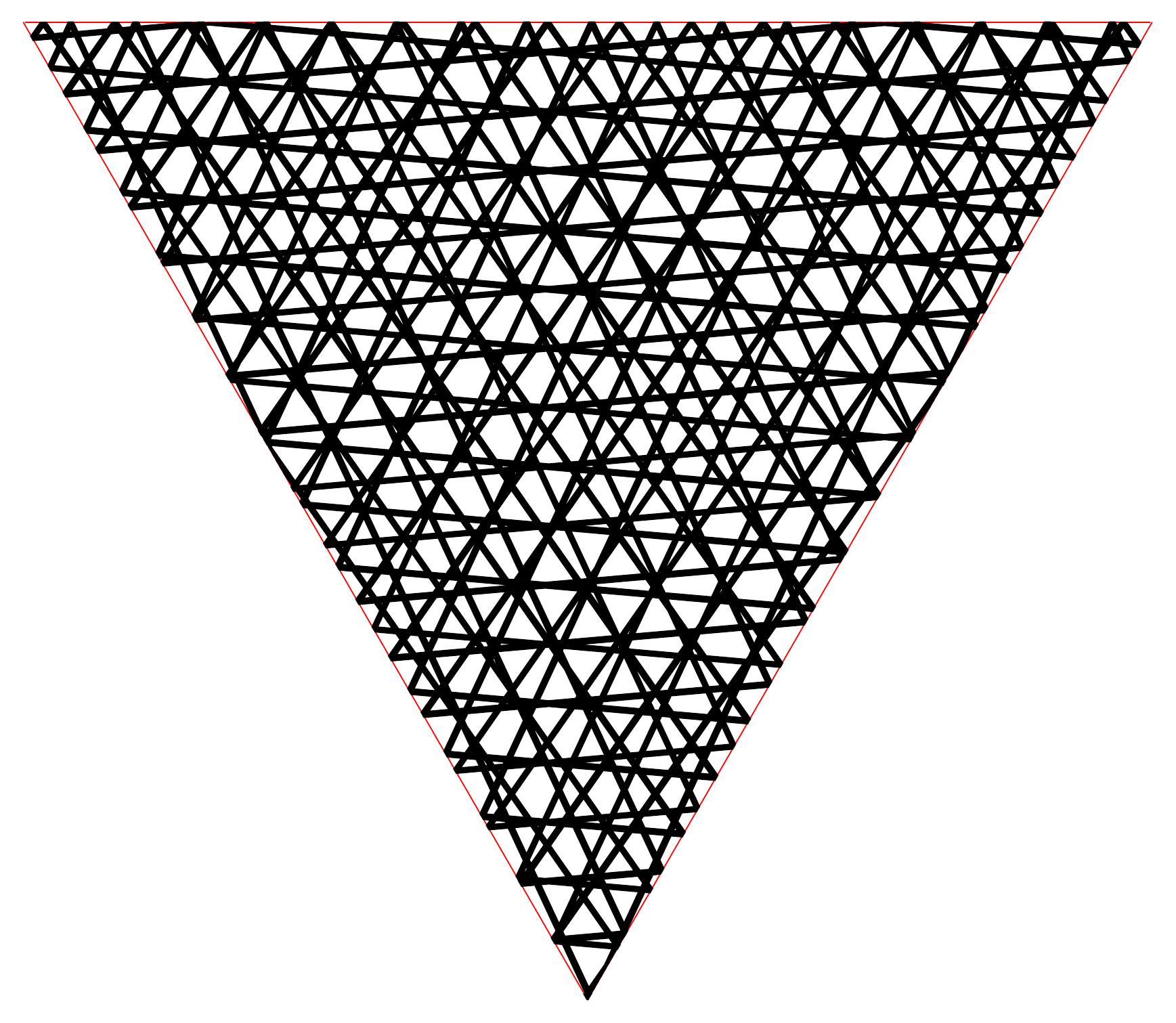}\\
  \includegraphics[width=0.1\textwidth]{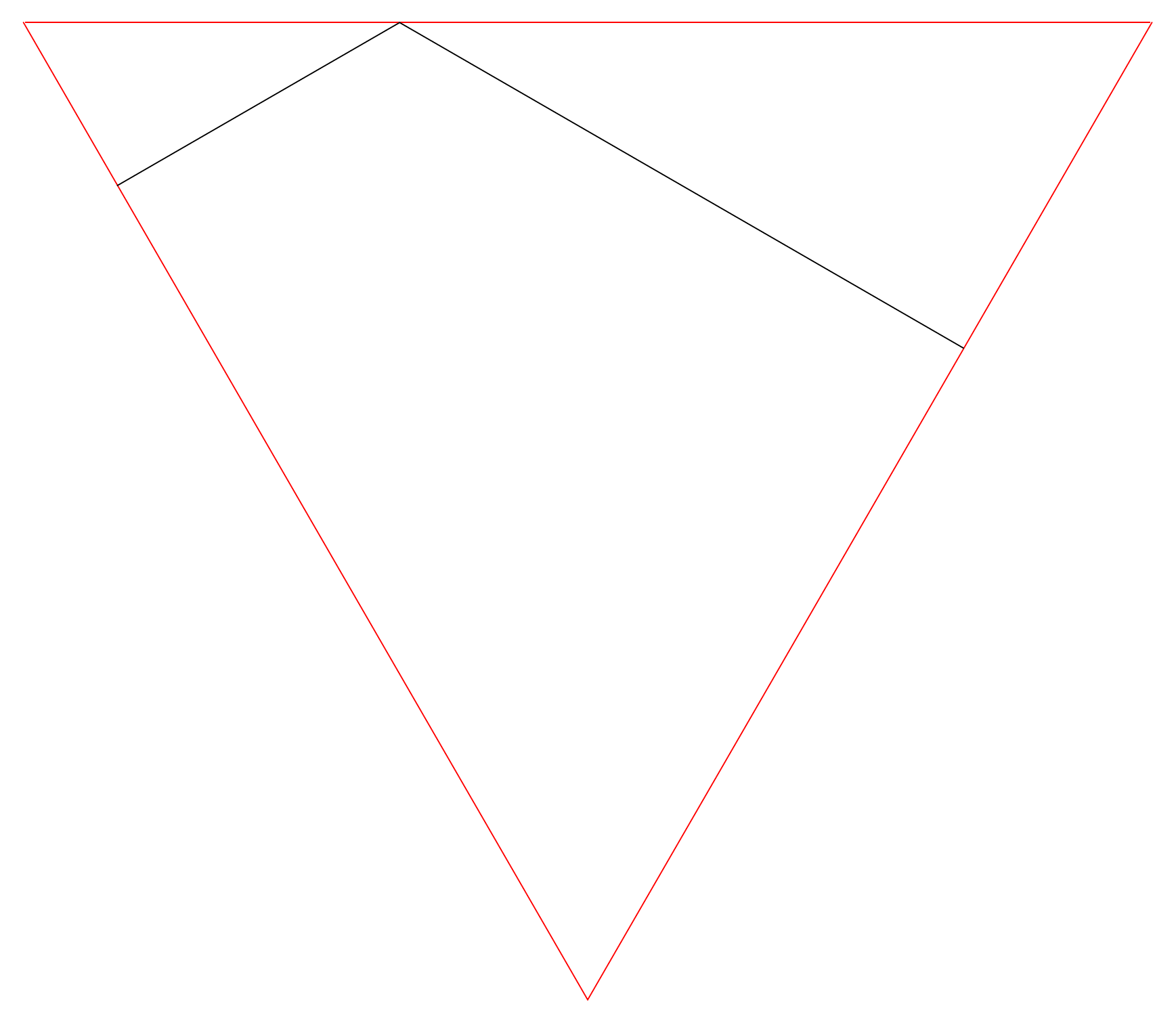}\hspace{0.3cm}
  \includegraphics[width=0.1\textwidth]{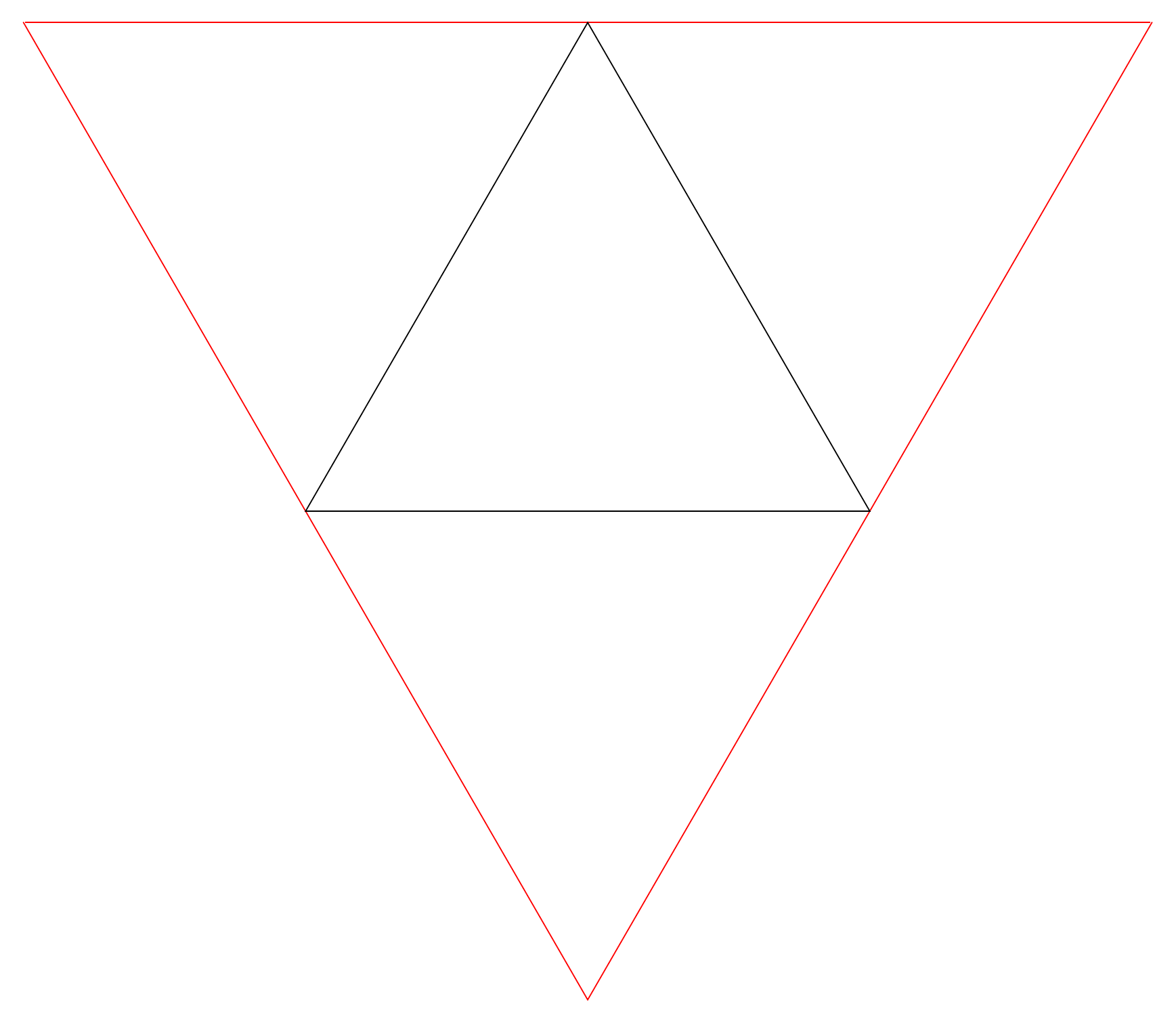}\hspace{0.3cm}
  \includegraphics[width=0.1\textwidth]{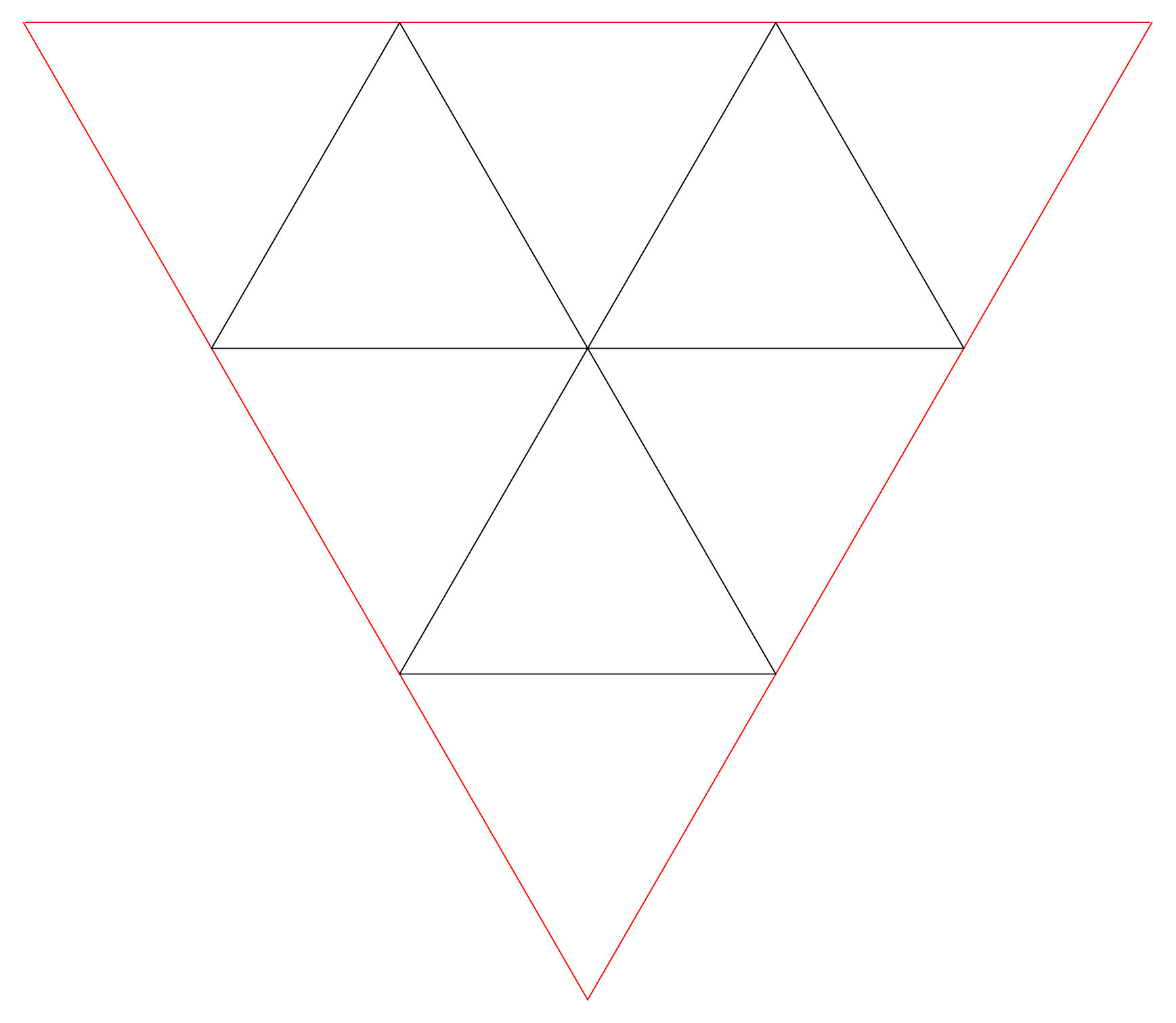}\hspace{0.3cm}
  \includegraphics[width=0.1\textwidth]{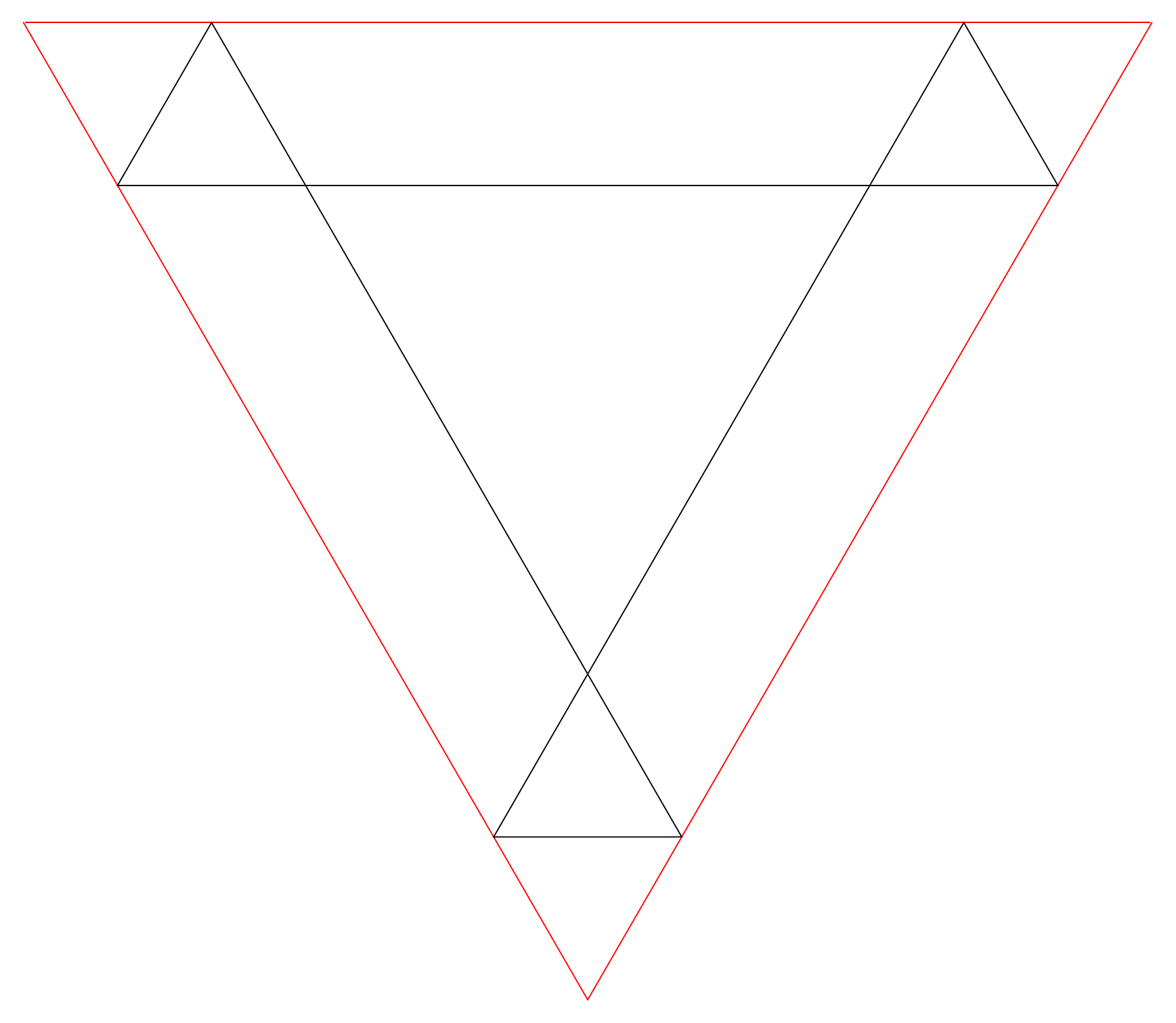}\\
  \includegraphics[width=0.125\textwidth]{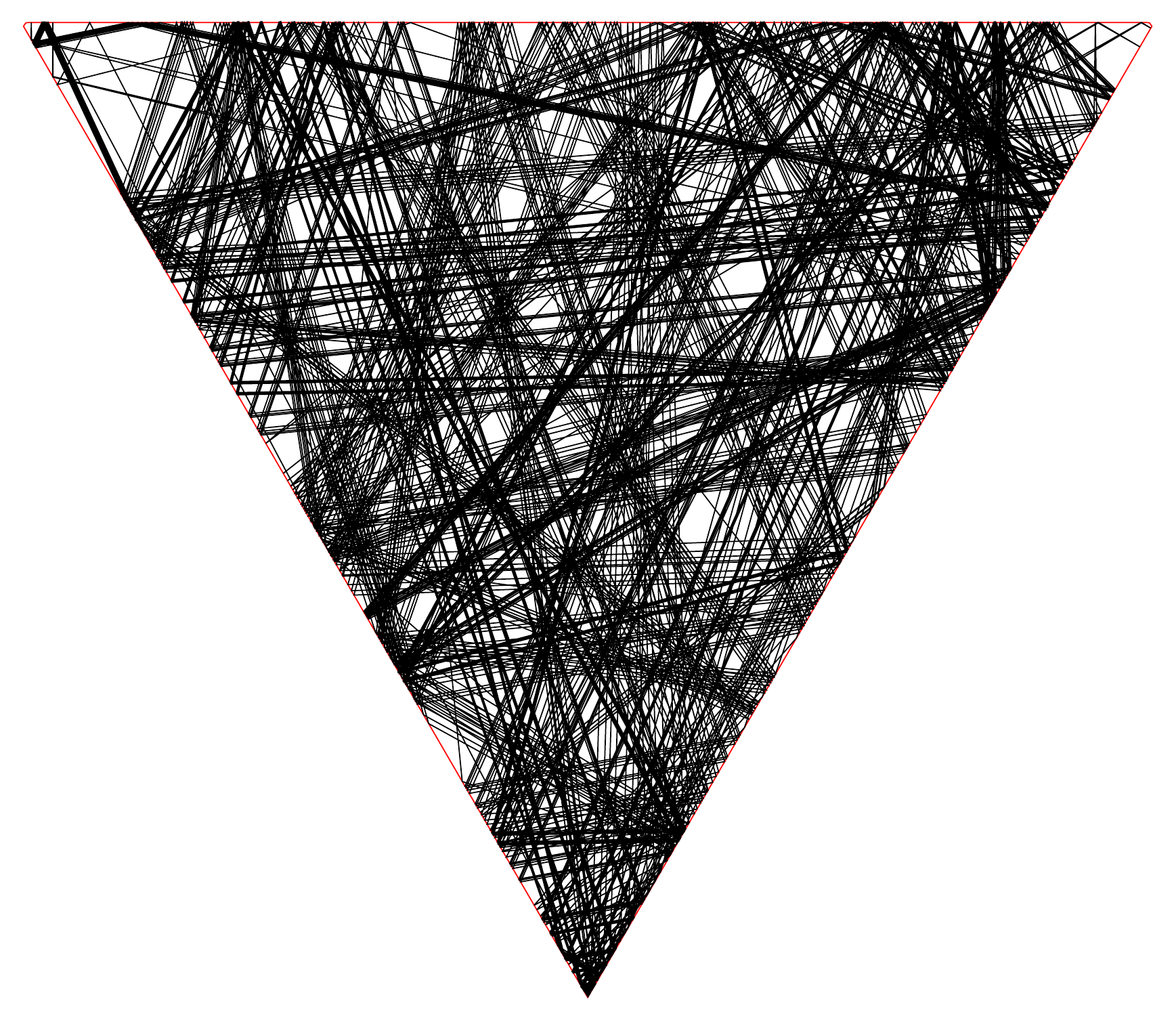}\hspace{1.0cm}    
  \includegraphics[width=0.15\textwidth]{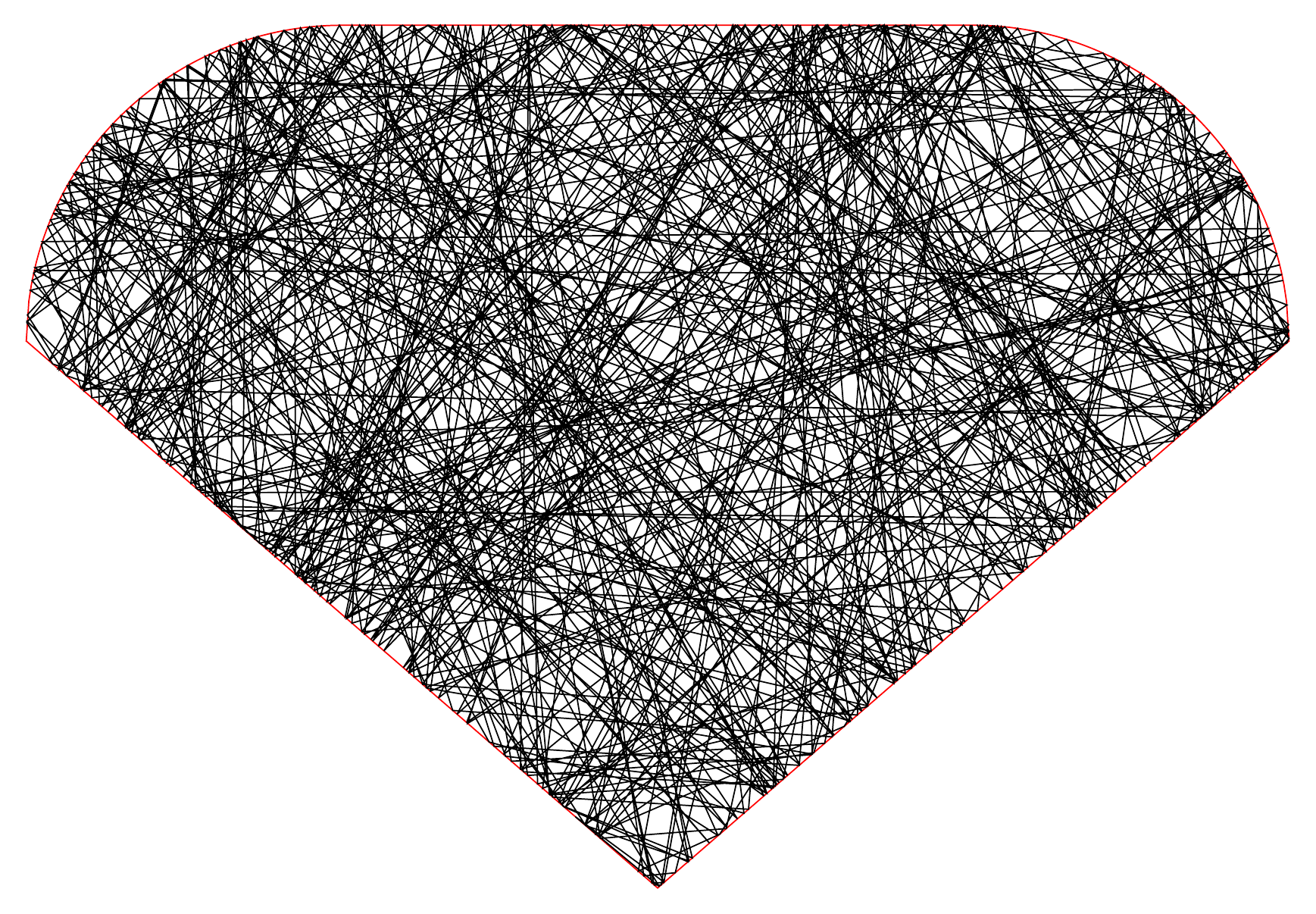}  
  			\caption[Classical trajectories in the hard wall billiard.]{
			\small Several trajectories after 1000 collisions are
            plotted. (\textit{Upper panel}) $\xi=1-1\times10^{-16}$, note that
            regular behaviour is obtained only for values of $\zeta$
            very close to 1. (\textit{Middle panel}) Some closed orbits of the
            triangular billiard. (\textit{Lower panel}) For $\xi$ far enough from 1 (diamond
            billiards), the trajectory fills the billiard irregularly
            (left) $\xi=0.99$ and (right) $\xi=0$.}
  \label{hardWallTrajectoriesFig}
\end{figure}
\\In general this procedure works well. Nonetheless, if the particle reaches one of the points $\left\{A,B,C,D,E\right\}$ where the tangent and normal vectors to the boundary are not defined, then the method fails. Although, this situation for an arbitrary initial condition rarely happens, the problem sometimes is solved by taking the average of the normal and tangent vectors for the boundaries connected in those problematic points. In Figure \ref{hardWallTrajectoriesFig} are shown some trajectories for the triangular billiard and the diamond billiard using the procedure described above. 

\subsubsection{Entropy}
In the previous section, a methodology based on geometry was used to
find the classical trajectories of the particle in a diamond
billiard. Indeed, this is not the more elegant way to find
trajectories, and there should exist a transformation or map which
connects the variables of the reduced phase space of consecutive
collisions of the diamond billiard. In principle, the trajectories of
the particle may be constructed with the knowledge of the billiard map and
the initial conditions. If we avoid the very special cases of the
periodic orbits, then the degree of irregularity of a set of
trajectories with different initial conditions should depend only on
the shape of the billiard. The entropy $S$ is calculated in order to
determine quantitatively such degree of irregularity. $S$ may be
computed as follows: Let $\alpha_n$ be the incident angle with respect
the normal vector on the boundary. The range of this variable is the
interval $I = [-\pi/2,\pi/2]$. This interval is divided in $M$ equal
subintervals $I_i$. Then $N$ collisions and their respective incident
angles $\alpha_n$ are generated, where $n=1,2,3,..,N$. If $N_i$ is the
number of angles $\alpha_n$ which live in the interval $I_i$, then the
probability to find an incident angle in the interval $I_i$ is
$P\left(I_i\right)=\frac{N_i}{N}$, and the entropy $S$ may be computed
in the standard way as
\begin{equation}
S = -\sum_{i=1}^N P\left(I_i\right)\ln\left[P\left(I_i\right)\right].
\label{entropyEq}
\end{equation}
\begin{figure}[h]
  \centering
  \includegraphics[width=0.43\textwidth]{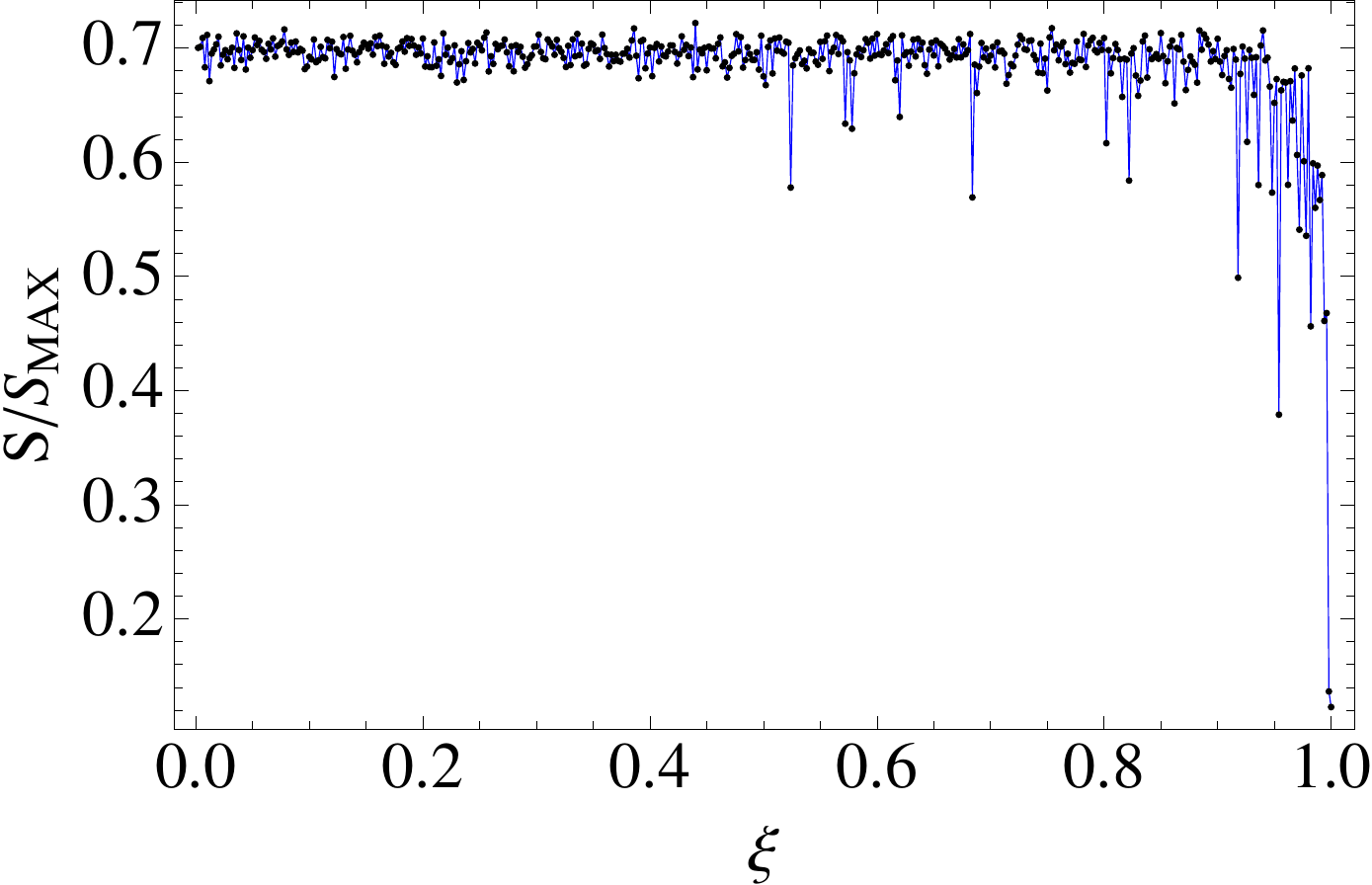}\hspace{0.1cm} 
  \includegraphics[width=0.43\textwidth]{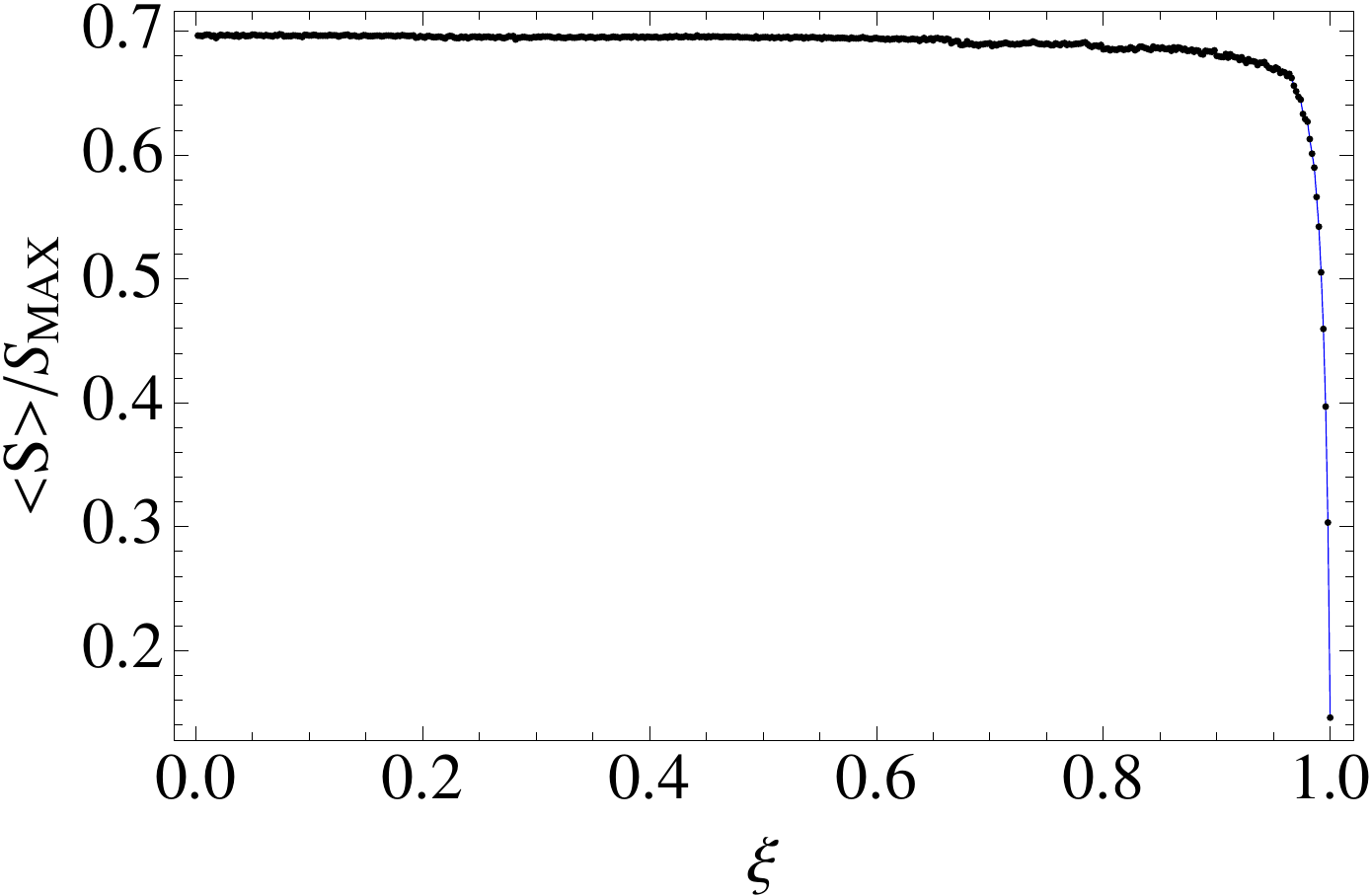}  
  \caption[Entropy]%
  {Entropy. (left) Entropy computed for a single value of $\xi$ and one initial random condition. The number of collisions of these graphs were $N=3000$, and the interval subdivisions were $M=100$. (right) We have generated 1000 random initial conditions with their respective entropies, and later these entropies were averaged.}
  \label{entropyFig}
\end{figure}
\\The maximum entropy is obtained when the set of generated incident angles $\left\{\alpha_n\right\}$ are uniformly distributed in the subintervals $\left\{I_i\right\}$. For this case the probability is equal for each subinterval, hence $P\left(I_i\right)=\frac{1}{N}$, and the entropy takes its maximum value $S_{max} = \ln(N)$. On the other hand, if all incident angles lie in a single
subinterval $I_j$, then the probability would be
$P\left(I_i\right)=\delta_{ij}$, and the entropy is zero. The entropy
computation using the equation (\ref{entropyEq}) generally depends on
the initial condition used. In order to avoid this dependence we have
computed the entropy for 1000 trajectories with different random
initial conditions. Posteriorly, these entropies were averaged for
each particular value of the control parameter (Figure
\ref{entropyFig}). The smallest value of entropy is obtained when
$\xi$ is exactly one and the billiard is an equilateral triangle. The
entropy grows quickly as the half of stadium is introduced in one of
the triangle sides, even when the control parameter is close to one as
$\xi=0.99$ where the corresponding entropy is
about the $55\%$ of its maximum theoretical value. As $\xi$ is set far from one, the entropy
practically stabilizes its value reaching about a $70\%$ of $S_{max}$.
In this regime, the trajectory of the particle is more complex than the one found for $\xi$ close to one as it is clear from a comparison between the lower and upper panels of Figure~\ref{hardWallTrajectoriesFig}.

\subsubsection{Lyapunov exponent}
The Lyapunov exponent $\lambda$ is used as a measure of divergence
between trajectories for a couple of infinitesimal close initial
conditions in the phase space. The time is not a suitable parameter in
order to compute $\lambda$ in billiard systems since the particle
movement is linear while it does not collide and the trajectories will
diverge linearly. The collision index $n$ was used as parameter
instead of time, as usual a great sensibility with small changes of
the initial conditions is characterized by $\delta_n = \delta_o
\exp\left(\lambda n\right)$ where $\delta_n$ is the absolute value of
the difference between incident angles of nearby trajectories after $n$
collisions.
\begin{figure}[h]
  \centering
  \includegraphics[width=0.35\textwidth]{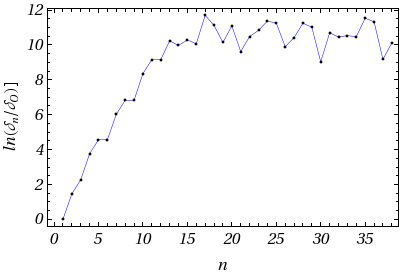} 
  \caption[Lyapunov exponent saturation]
  {Lyapunov exponent saturation. This is a typical graphic obtained for a single
    value of the control parameter ($\xi=0.9999$) and one random
    initial condition; 40 collisions are considered. For this
    particular situation the saturation occurs about after 13
    collisions. The Lyapunov exponent is the slope of the
    non-saturated part.}
  \label{deltaGraphicFig}
\end{figure}\\
\begin{figure}[h]
  \centering
  \includegraphics[width=0.8\textwidth]{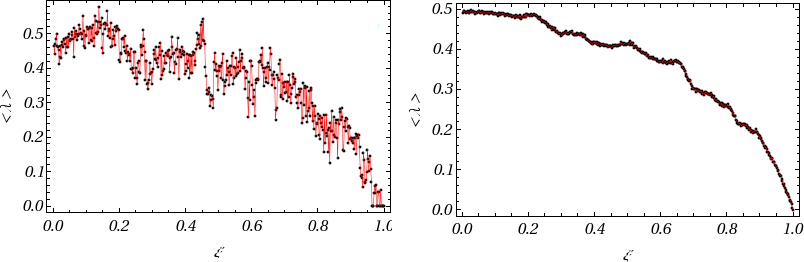} 
  \caption[Averaged Lyapunov exponent.]%
  {Averaged Lyapunov exponent. The number of random initial conditions of each graph was
    (left) 10 and (right) 1000.}
  \label{lyapunovFig}
\end{figure}
The Lyapunov exponent was averaged in order to avoid the dependence with the initial conditions. For a couple of close trajectories random initial conditions were generated, and the Lyapunov exponent was computed for each initial condition (Figure \ref{deltaGraphicFig}). The Lyapunov exponents computed in the previous step were averaged for each value of $\xi$ (Figure \ref{lyapunovFig}). In order to minimize the error introduced by the saturation, we decided to calculate the slope between adjacent points and average it for each single Lyapunov exponent computed, thus the saturated points frequently have small contribution due to the alternation of the slope sign. Some graphics are not well defined as the one shown in Figure \ref{lyapunovFig} and some inaccuracies persist in the final result, even if the number of initial conditions is increased. For this reason, the final averaged on the Lyapunov exponent in Figure \ref{lyapunovFig} is not as smooth as the entropy of the Figure \ref{deltaGraphicFig}. However, the Figure \ref{lyapunovFig} is able to capture an important feature of the billiard: as the half stadium appears over one side of the original equilateral triangle, then the Lyapunov exponent substantially increases, and the non negative values of it ensures a great sensibility to the initial conditions, even for values of $\xi$ close to one.

\section{Quantum diamond billiard: Finite Difference Method Implementation}
\begin{figure}[h]
  \centering
  \includegraphics[width=0.27\textwidth]{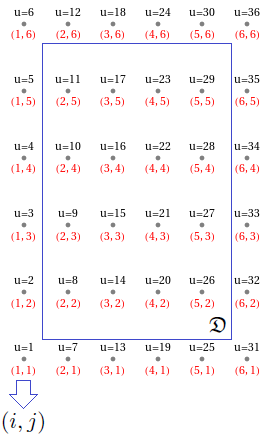}\hspace{0.1cm}
  \includegraphics[width=0.47\textwidth]{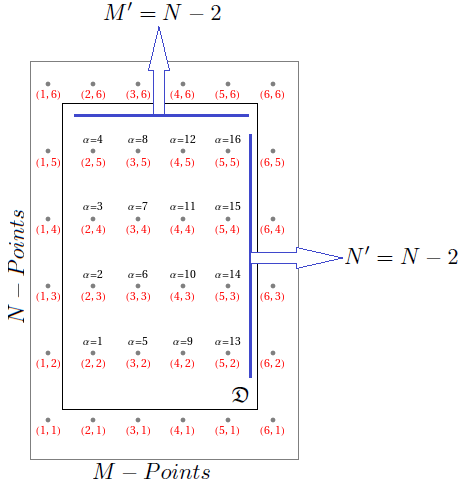} 
  \caption[First indexing]%
  {First and second indexing. (left) The domain is
    discretized in a rectangular lattice of $M\times N$ points
    including the boundary (the points out the solid rectangle). Each
    point $(i,j)$ is labelled by the single index $u$. (right) Here
    only the inner points are indexed, we build the map
    $\alpha=\alpha(i,j)$, inversely the pair $(i,j)$ is obtained by
    $i=i_2(\alpha)$ and $j=j_2(\alpha)$. The points of the two
    indexing are related by
    $u=u(i,j)=u\left(i_2(\alpha),j_2(\alpha)\right)$. The total number
    of inner points is $Q=M'\times N'$ }
  \label{firstIndexing}
\end{figure}
\begin{figure}[h]
  \centering
  \includegraphics[scale=0.25]{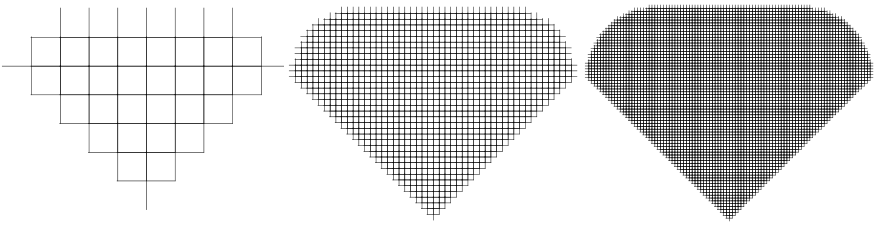} 
  \caption[The grid.]%
  {The grid. As the density of points is incremented, the mesh is better adjusted to the geometry of the billiard. Some grids are shown, the number of inner points are: (left) $Q=32$, (center) $Q=1108$ and (right) $Q=4618$.}
  \label{gridFig}
\end{figure}
\begin{figure}[h]
  \centering
  \includegraphics[width=0.33\textwidth]{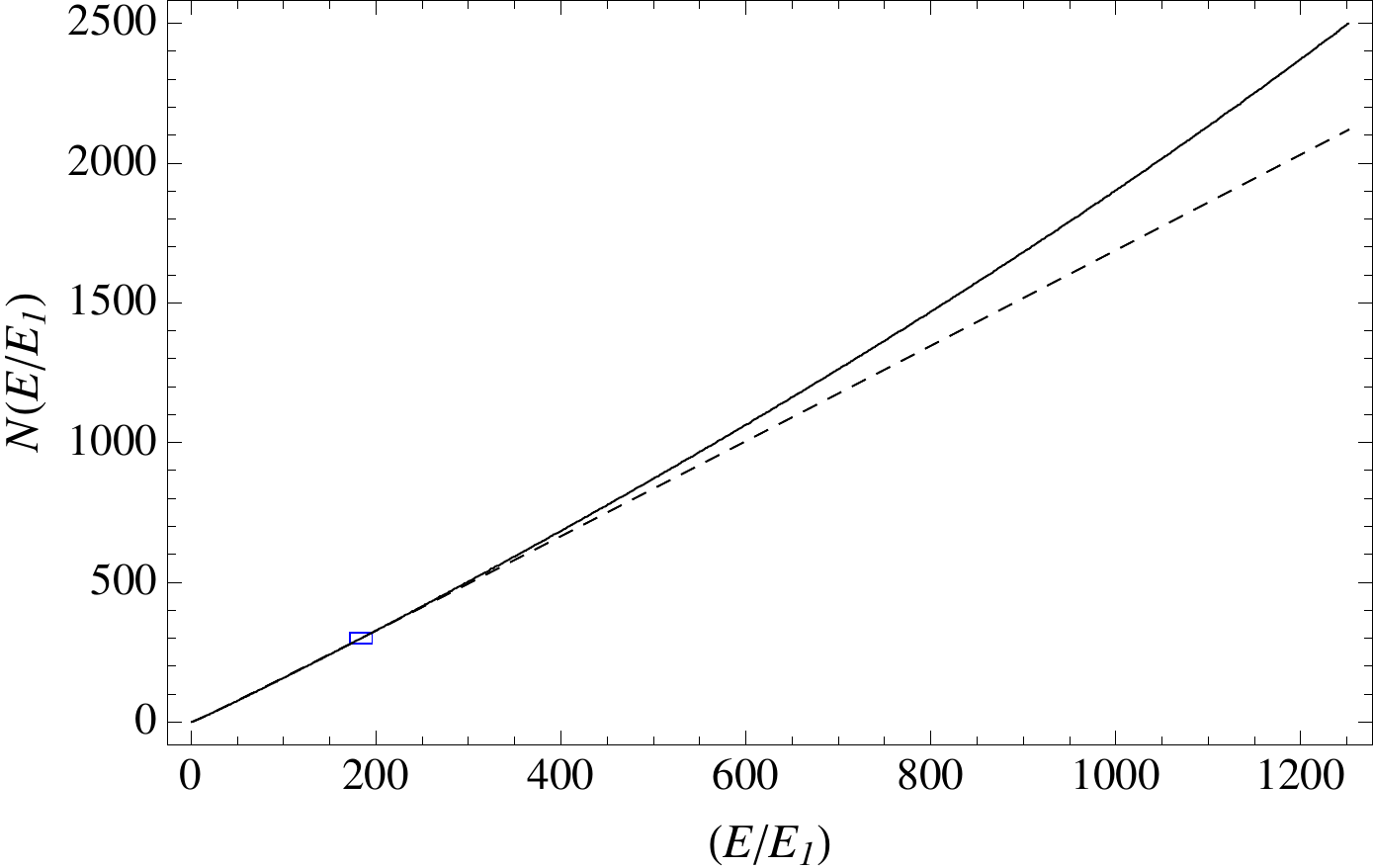}\hspace{0.05cm}
  \includegraphics[width=0.33\textwidth]{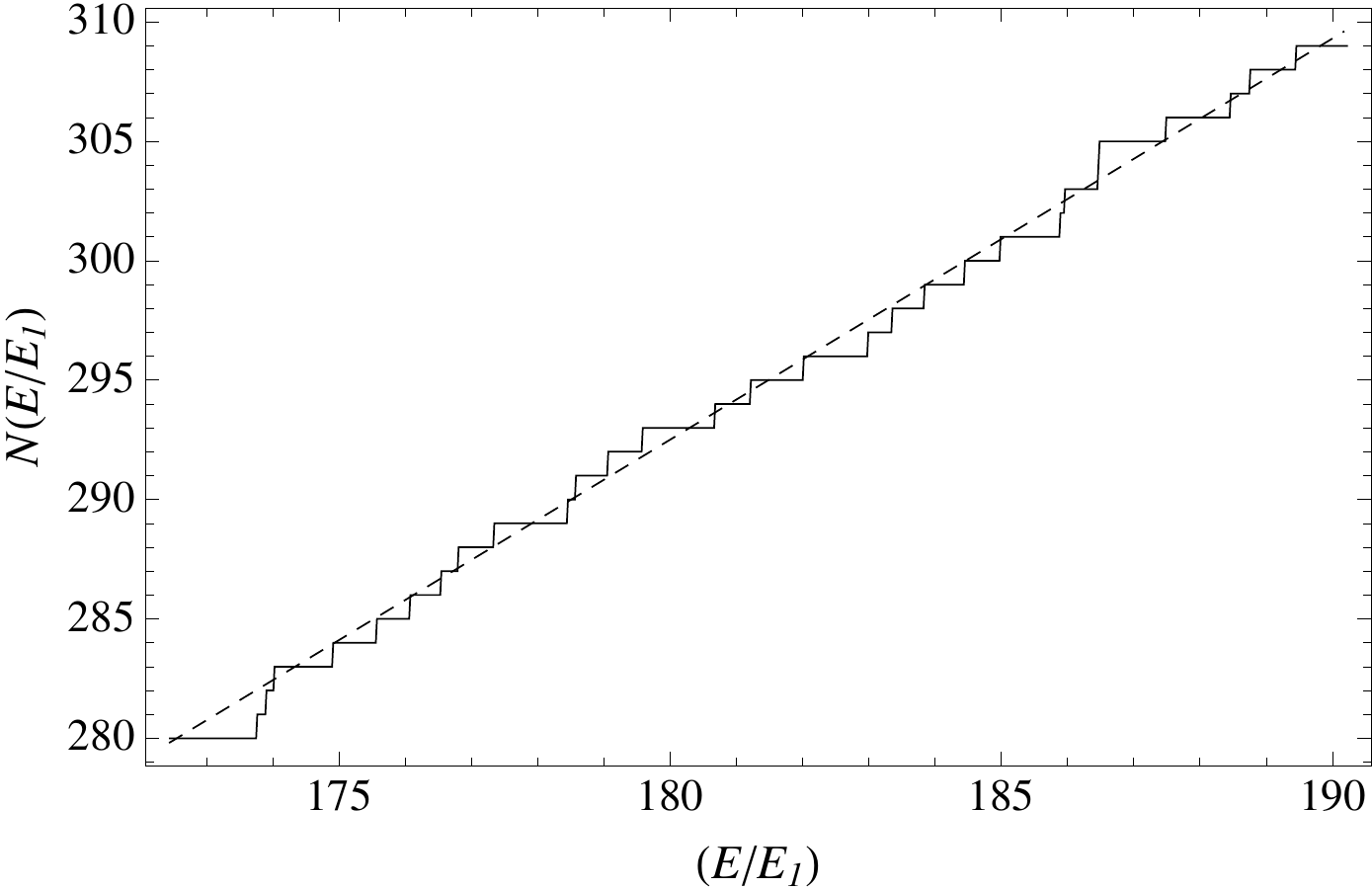} 
  \caption[Spectral staircase function of the diamond shaped
    billiard.]
  {Spectral staircase function for the diamond shaped billiard. (left) Superposition of the numerical staircase energy staircase function (solid line), with the Weyl's formula (dashed line). (right) A zoomed region corresponding to the blue rectangle of the graphic on the left side.}
  \label{weylProofCompletedBilliardFig}
\end{figure}

This problem is typically solved by using \textit{finite element method} (FEM) and iterative
methods on the discretized version of the Schr\"odinger equation
\cite{scarringInOpenQC}. We may use the FDM
to express the Hamiltonian as a matrix on a lattice, and then
solve the resulting eigenvalue problem. The discretization of the region is shown in Figure
\ref{firstIndexing}-left, each point at the position $\vec{r}_{ij}$ is
labelled in one of these ways: with pair $(i,j)$ or the single index
$u$. The second option is used in order to avoid the impractical use
of four indices in the Hamiltonian matrix. During the lattice
construction it is easy to build the function $u=u(i,j)$ which maps
from the pair of indices $(i,j)$ to the point $u$, let us call this
process as the \textit{first indexing}. The time independent
Schr\"odinger equation is evaluated at the point $u(i,j)$ according to
\begin{equation}
- \left.\frac{\hbar}{2m} \vec{\nabla}^2\psi(\vec{r})\right|_{u(i,j)} + \left.V(\vec{r})\psi(\vec{r})\right|_{u(i,j)} = E \left.\psi(\vec{r})\right|_{u(i,j)}. 
\label{eqTDSEevaluatedInU} 
\end{equation}
The second derivatives of the laplacian may be evaluated using
central differences~\cite{fdmInElectromagnetism}
\[
\left.\frac{\partial^2\psi}{\partial x^2}\right|_{u(i,j)} \approx \frac{1}{\delta x^2}\left[\psi_{u(i+1,j)}+\psi_{u(i-1,j)}-2\psi_{u(i,j)}\right]\] 
and
\[\left.\frac{\partial^2\psi}{\partial y^2}\right|_{u(i,j)} \approx \frac{1}{\delta y^2}\left[\psi_{u(i,j+1)}+\psi_{u(i,j-1)}-2\psi_{u(i,j)}\right]. 
\]
The notation is simplified defining
\begin{equation}
u^\pm = u\left(i(u)\pm 1,j(u)\right).
\label{horizontalDisplacement}
\end{equation}
This transformation makes a horizontal displacement in the lattice
from the point $u$. Although, there are several values of $u$ for a
single $i$ or $j$ ($N$ values for the index $i(u)$, and $M$ values for
the index $j(u)$) we have only one value for $u$ fixing both $i$ and
$j$. The equation (\ref{horizontalDisplacement}) gives the neighbor of
$u$ at its left $(-)$ or right $(+)$.  Similarly, the vertical
displacement from the point $u(i,j)$ is computed with
\begin{equation}
u_\pm = u\left(i(u),j(u)\pm 1\right) \,.
\end{equation}
With this notation, equation (\ref{eqTDSEevaluatedInU}) may be written as 
\[
\sum_{v\in\mathfrak{D}}H_{uv}\psi_v = E \psi_u \hspace{1.0cm}\mbox{where}
\] 

\begin{equation}
H_{uv} =
-\frac{\hbar^2}{2m}\left[\frac{\delta_{u^+,v}+\delta_{u^-,v}}{\delta
    x^2}+\frac{\delta_{u_+,v}+\delta_{u_-,v}}{\delta
    y^2}-2\delta_{u,v}(\delta x^{-2}+\delta
  y^{-2})\right]+V_u\delta_{u,v}
\label{HamiltonianMatrix}
\end{equation}
is the Hamiltonian (repeated indices in the last term do not involve sum over them). Since the problem is solved only for the inner points, we performed a
\textit{second indexing} (see Figure \ref{firstIndexing}-right), so
the eigenvalue equation may be written as
\begin{equation}
\sum_{v\in\mathfrak{D}}H_{uv}\psi_v = \sum_{\beta=1}^{Q=M'\times N'}H_{u(\alpha)v(\beta)}\psi_{v(\beta)} = E \psi_{u(\alpha)} .
\end{equation}
The eigenvalues and eigenvectors of the Hamiltonian $H_{\alpha\beta}$ are obtained numerically. Commonly, the packages of matrix diagonalization arrange the eigenvectors in a matrix, let us call it $M_{\alpha \beta}$\\  
\begin{equation}
M =
\left( {\begin{array}{*{20}c}
		M_{11} & M_{12} & \cdots & M_{1Q}\\
		M_{21} & M_{22} & \cdots & M_{2Q}\\
  		\vdots & \vdots & \vdots & \vdots \\
  		M_{Q1} & M_{Q2} & \cdots & M_{QQ}
 \end{array} } \right) .
\end{equation}
If the eigenvectors are arranged in the columns of such matrix, then the state $s$ evaluated at the point $\alpha$ is $\psi_\alpha^{(s)} = M_{\alpha s}$ with $(s=1,2,...,Q)$. In order to return to the initial labelling we may write
\begin{equation}
\psi_{ij}^{(s)} = \left\{ \begin{array}{rl}
 M_{\alpha(i,j),s} &\hspace{0.5cm} \mathrm{if} \hspace{0.5cm} \vec{r}_{ij}  \in \mathfrak{D} \\
  0 & \hspace{0.5cm} \mathrm{otherwise.} 
       \end{array} \right. 
\end{equation}
The generalization for billiards of arbitrary shape does not
represent considerable difficulties. We may place the boundary of the
arbitrary shape billiard over the rectangular grid and take only the
points inside of it. After the identification of the boundary points,
the inner points (say $Q$ inner points) are enumerated first
$(u=1,2,...,Q)$, and the boundary points later, so a second indexation
is avoided. Some grids for the billiard consider in this study are
shown in Figure \ref{gridFig}.
\\
\begin{table}
\begin{center}
\caption{Triangular billiard eigenvalues.}

\begin{tabular}{l*{3}{c}}
State             & $E_n$ (J) & $E_n$ (J) & Relative Error \\
$n$               & \textbf{FDM} & \textbf{Exact} & $\%$ \\
\hline
\hline
1 & 2.14940 & 2.14849 & 0.04235   \\
2 & 5.01655 & 5.01313 & 0.06817   \\
3 & 5.01743 & 5.01313 & 0.08570   \\
4 & 8.59995 & 8.59394 & 0.06988   \\
5 & 9.31609 & 9.31010 & 0.06429   \\
6 & 9.31662 & 9.31010 & 0.06998   \\
7 & 13.6095 & 13.6071 & 0.01763   \\
8 & 13.6171 & 13.6071 & 0.07343   \\
9 & 15.0451 & 15.0394 & 0.03788   \\
10 & 15.0451 & 15.0394 & 0.03788  \\
11 & 19.3386 & 19.3364 & 0.01137  \\
12 & 20.0559 & 20.0525 & 0.01695  \\
13 & 20.0559 & 20.0525 & 0.01695   \\
14 & 22.1997 & 22.2010 & 0.00585   \\
15 & 22.2001 & 22.2010 & 0.00405   \\
\end{tabular}
\end{center}
\label{energyLevelsTable}
\end{table}
The numerical results were compared with the exact ones for the
equilateral triangular billiard. The triangular billiard is integrable
and the expression for the energy levels is well known
\cite{SemmiclassicalPhys}
\begin{equation}
E_n = E_{pq} = \left(\frac{4\pi}{3}\right)^2\left(\frac{\hbar^2}{2md_1^2}\right)\left(p^2+q^2-pq\right)
\label{triangleLevelsEq}
\end{equation}
where $d_1$ is the edge length when $\xi=1$,
$p$ and $q$ are positive integers which satisfy $q\in\left[1,p/2\right]$. A
comparison of the numerical and analytic energy levels was done and
listed in Table 1. 
One advantage of the FDM lies in the fact that Hamiltonian is computed by a direct
evaluation of the potential and some Kronecker deltas, so in a
personal computer (in our case an i7 processor) building a Hamiltonian
matrix of $11000\times11000$ take less than a minute, and its
orthogonalization with the lapack package using Fortran, requires
about 25 minutes. In order to check the accuracy of the results, a
comparison with the \textit{energy staircase function}
$\mathcal{N}(E)$ with the Weyl-type formula was
performed. $\mathcal{N}(E)$ gives the number of energy levels under
the energy $E$ and it is defined by
\begin{equation}
\mathcal{N}(E) := \sum_i \theta\left(E-E_i\right)
\end{equation}\\
where $\theta(x)$ is the step function. The analytical result for a two-dimensional billiard with area $A$ and perimeter $P$ is given by \cite{weyl1,weyl2,MKac,circularBilliard}
\begin{equation}
\mathcal{N}(E) = \frac{A}{4\pi}\left(\frac{2mE}{\hbar^2}\right)-\frac{P}{4\pi}\sqrt{\frac{2mE}{\hbar^2}}+o(E^{1/2}).
\label{weylFormula}
\end{equation}\\
Carefully speaking, the Weyl formula is valid in
the semiclassical limit, that is for high energy levels. 
\begin{figure}[h]
  \centering
  \includegraphics[width=0.15\textwidth]{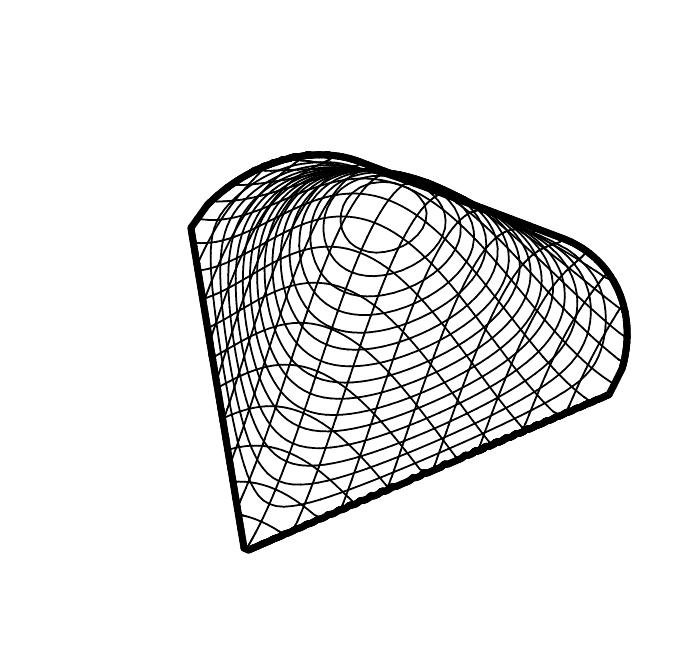}    
  \includegraphics[width=0.15\textwidth]{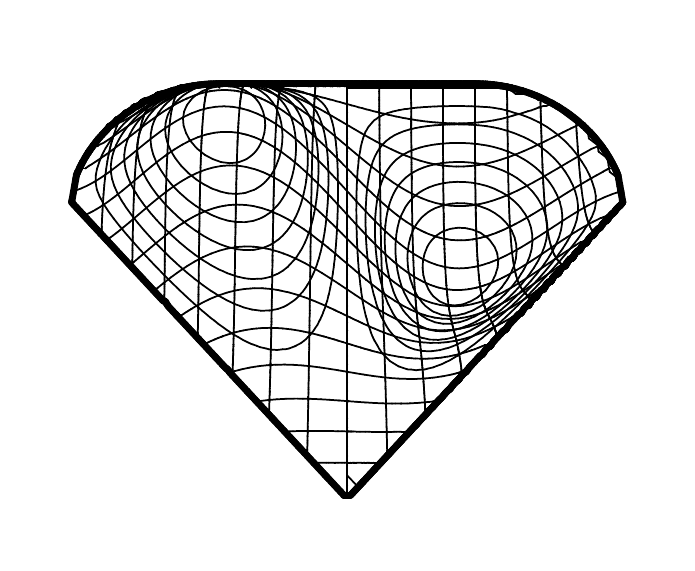}    
  \includegraphics[width=0.15\textwidth]{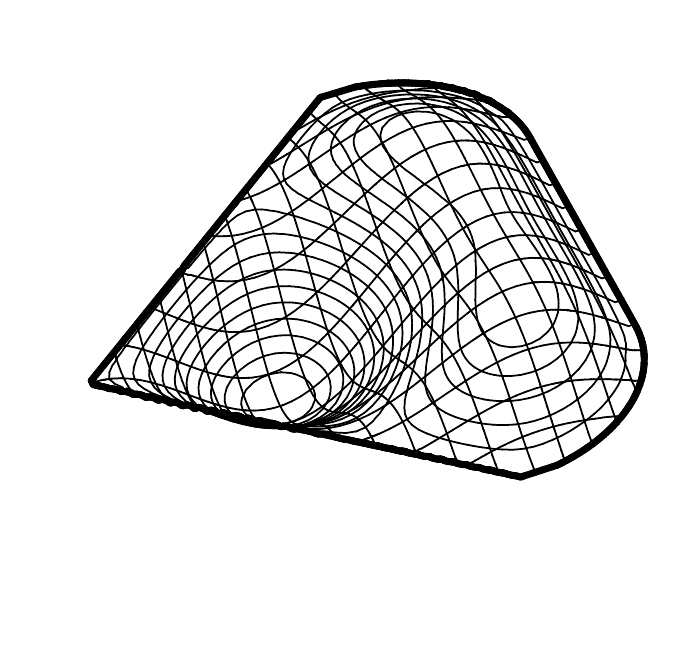}   
  \includegraphics[width=0.15\textwidth]{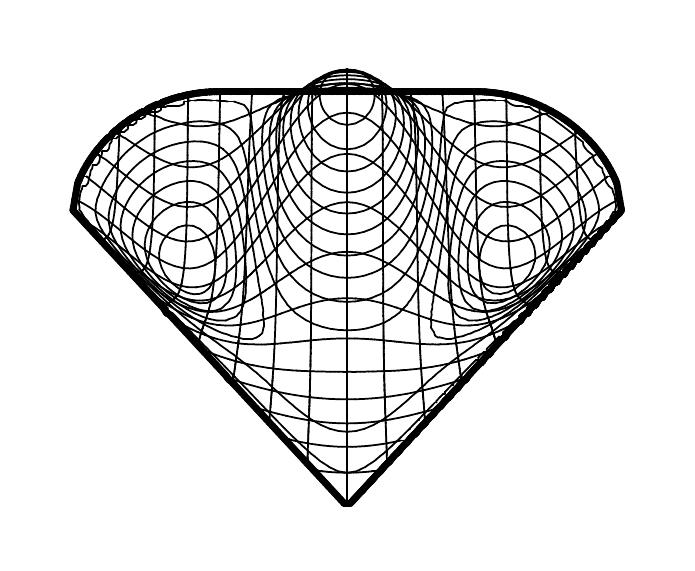}    
  \includegraphics[width=0.15\textwidth]{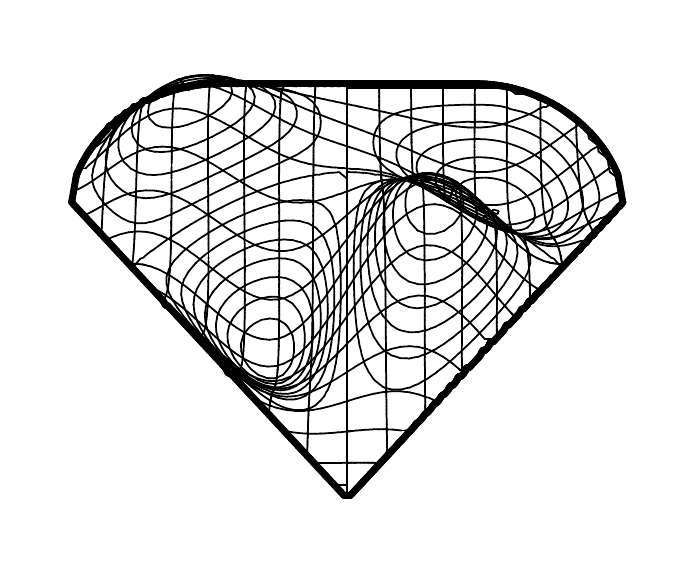}\\    
  \includegraphics[width=0.15\textwidth]{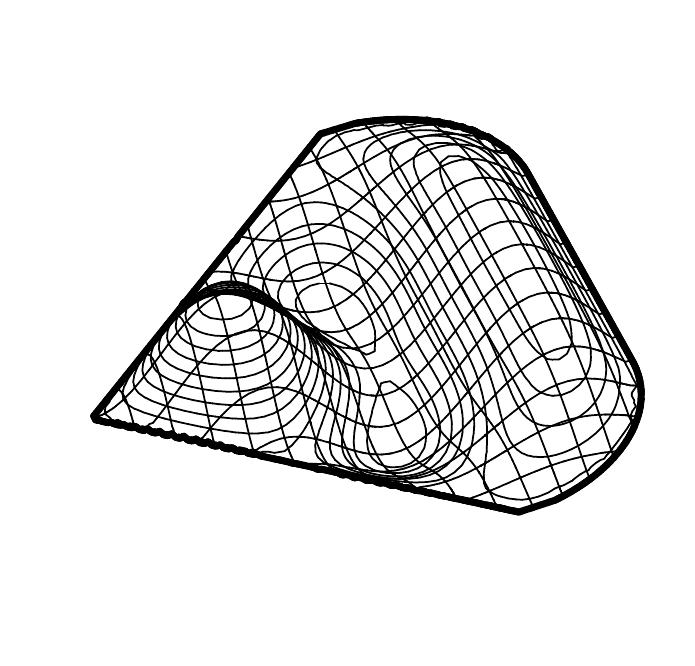}    
  \includegraphics[width=0.15\textwidth]{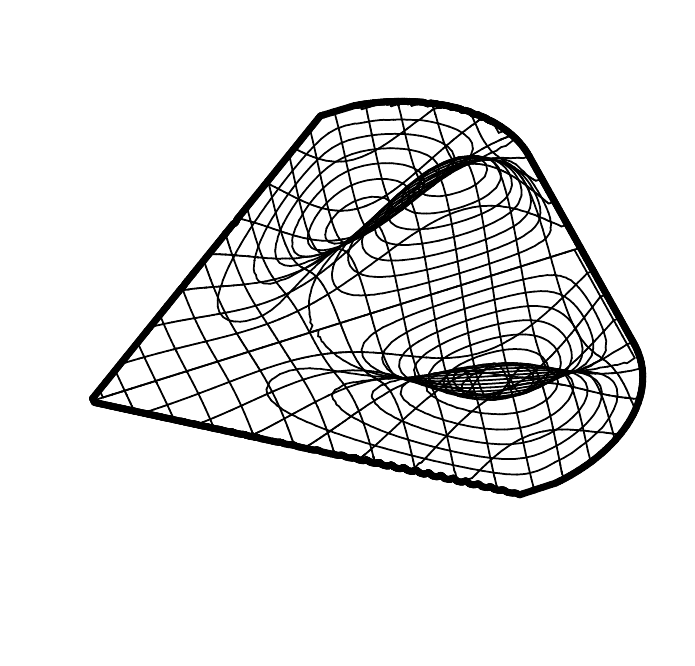}    
  \includegraphics[width=0.15\textwidth]{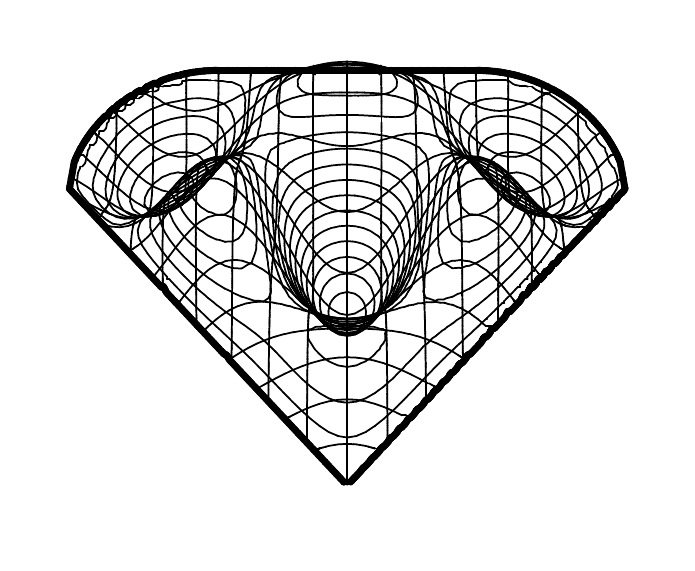}    
  \includegraphics[width=0.15\textwidth]{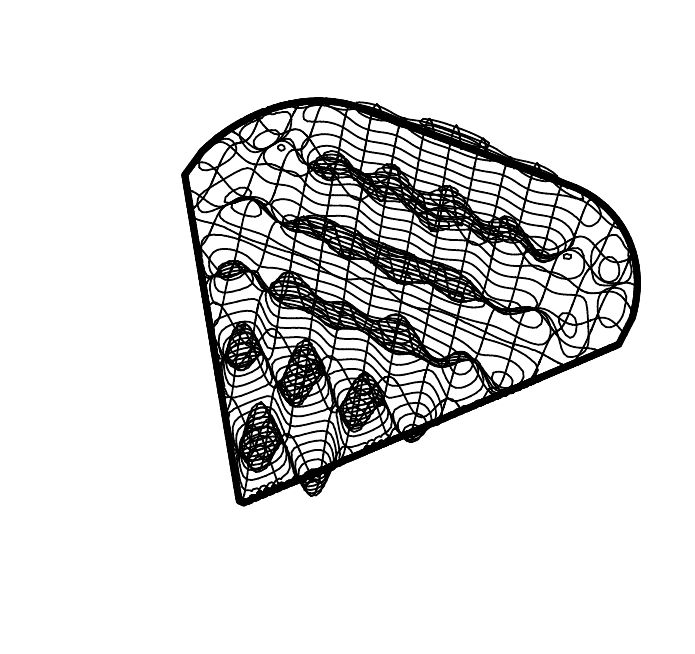}   
  \includegraphics[width=0.15\textwidth]{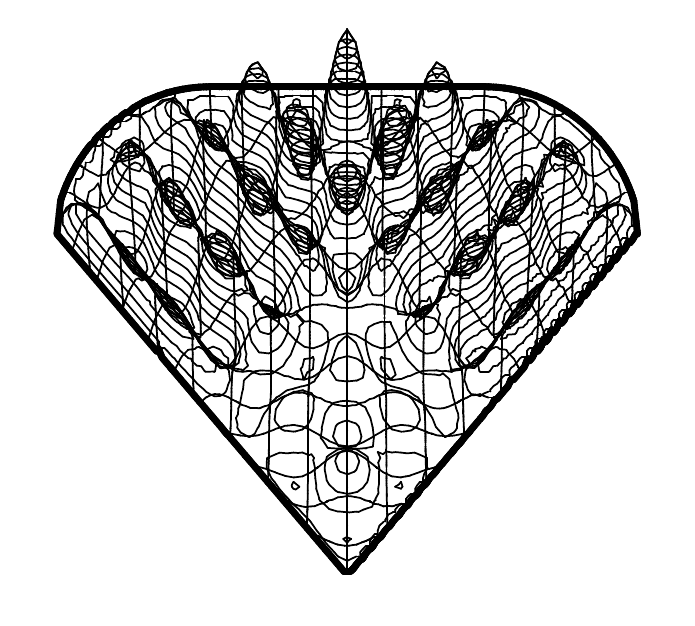}   
  \caption[Diamond shape billiard wavefunctions.]%
  {Diamond shape billiard wavefunctions. Here some eigenfunctions are plotted using the a grid with $Q=4618$ inner points (see Figure \ref{gridFig}), and the control parameter was set as $\xi=1$. We have set $m=1$ and $\hbar=1$.}
  \label{statesFig}
\end{figure}
Other typical problems in billiards as the scar identification deep in
the semiclassical limit may be faced using more convenient
methods. For instance, the \textit{improved Heller's PWDM (Heller's
  plane wave decomposition method)} \cite{PWDM}. This method avoids
the computation of the eigenvectors near to the ground state and goes
directly for the computation of the states with high quantum numbers. We used the finite difference method in order to diagonalize quantum
diamond billiard. Some of the first excited states of this billiard
are shown in Figure \ref{statesFig}. In Figure~\ref{weylProofCompletedBilliardFig} we have superposed the
numerical result of the energy staircase function with the one
provided by~(\ref{weylFormula}). The deviation of the numerical
results to the theoretical prediction for high energies is due to
numerical errors because the discretization procedure is not able to
describe properly wavefunctions corresponding to very high energies,
which have very small wavelength oscillations. From this, it is clear
that only the first $\sim 200$ computed states are reliable.\\

\section{Quantum diamond billiard: level statistics}
\begin{figure}[h]
  \centering    
  \includegraphics[width=0.35\textwidth]{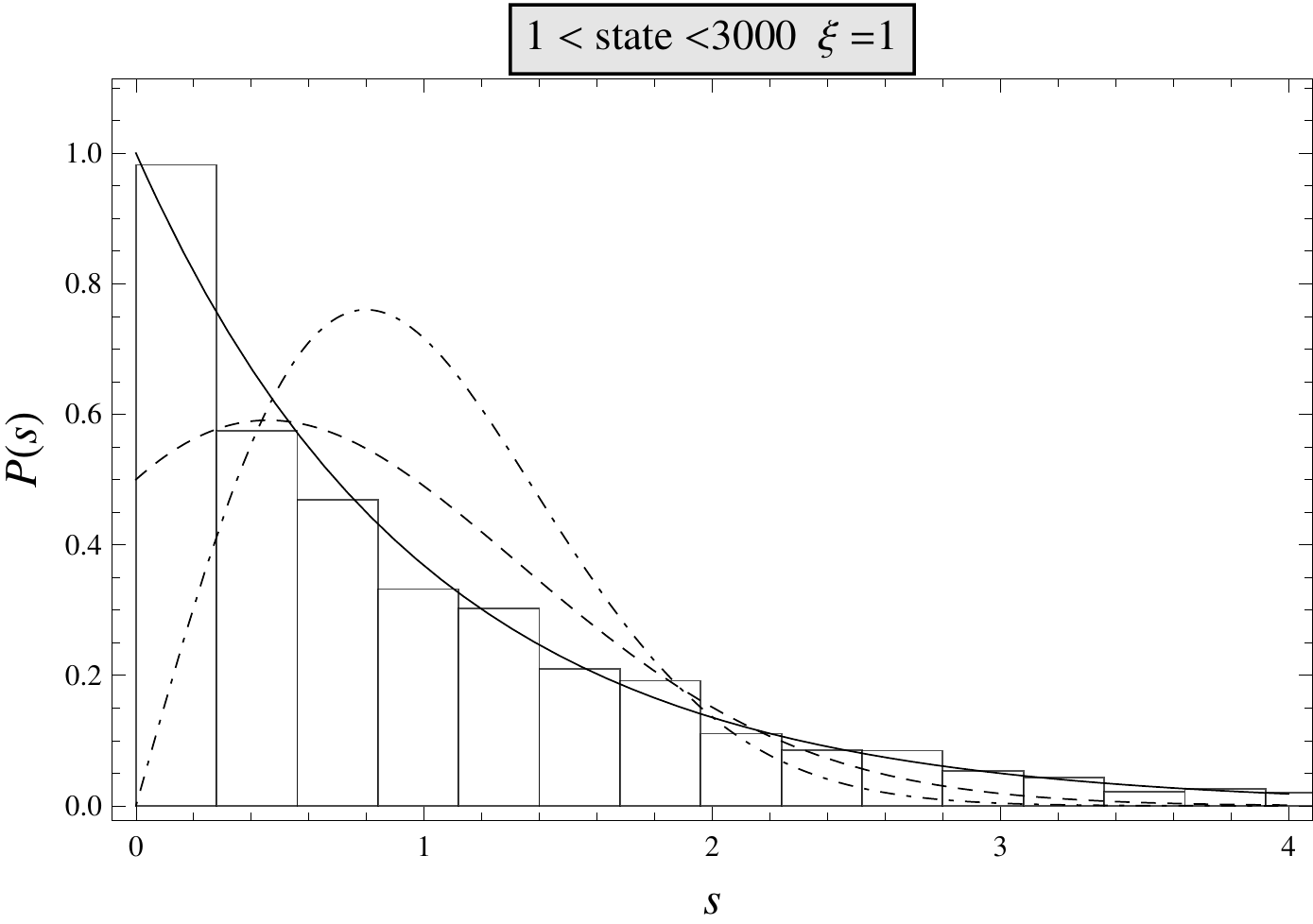}  
  \caption[NNSD of the triangular billiard.] {Nearest
  neighbor spacing distribution of the energy levels of the triangular
  billiard. The histogram was built using the analytic expression for
    the energy levels of the equilateral triangular billiard (see
    equation (\ref{triangleLevelsEq})). The level statistics is
    Poissonian (solid line) because the equilateral triangle billiard
    is regular. For the irregular case, the nearest neighbor spacing
    distribution will follow either a GOE2 distribution (dashed line)
    or a GOE distribution (dot dashed line) according to the billiard
    symmetries.}
  \label{poissonFig}
\end{figure}

Several experiments were performed in the nineties with quantum hard
wall billiards, e.g. the microwave resonators, which used the
equivalence between the stationary Scr\"odinger equation with the Helmholtz
equation to study chaos in quantum billiards using electromagnetic
waves \cite{microWaveResonators}.  Other devices used in the quantum
chaos study were the semiconductors billiards, those are open quantum
cavities which permit a current flow through two contact
points. These structures are different from a quantum billiard, which
is completely closed and confines a single particle inside
it. However, if the size of the quantum open cavity is much smaller
than the mean free path of the electrons, then the device approaches a
quantum billiard and it shares with the quantum billiards several
properties e.g. energy level statistics and the scarring of the
wavefunction \cite{scarringInOpenQC}. For Hamiltonian systems such as
the one described in this writing the energy level spacing
distribution, $P(s)$, is a feature which distinguishes the spectrum of a
system with regular or chaotic classical analogue. According to the
\textit{Bohigas-Giannoni-Schmit} conjecture \cite{bgsCite} the spectra
of a time-reversal-invariant system with a classical chaotic
counterpart follows a Gaussian Orthogonal Ensamble (GOE)
distribution. On the other hand, if the classical analogue is regular,
then the spectrum is characterized by a Poisson distribution (see
Figure \ref{poissonFig}). This conjecture has been tested in a variety
of systems including billiards.

The diamond shaped billiard has a mirror reflection symmetry axis. For
this reason, there are two set of states related to each symmetry
class, namely, odd or even eigenstates. The general expression of the
nearest neighbour spacing distribution for a superposition of $N$
independent spectra in the GOE statistics is given by~\cite{GeneralGoe}
\begin{equation}
P_{N\,\mathrm{GOE}}(s) = \frac{\partial^2}{\partial s^2}\left[\erfc\left(\frac{\sqrt{\pi} s}{2N}\right)\right]^N 
\end{equation}
where $s$ is the energy spacing between nearest neighbour levels, and  $\erfc(x)$ is the complementary error function. The spacing distribution for $N=2$ is
\begin{equation}
P_{2\,\mathrm{GOE}}(s) = \frac{1}{2}e^{-\frac{s^2\pi}{8}} + \frac{\pi s}{8}e^{-\frac{s^2\pi}{10}}\erfc\left(\frac{\sqrt{\pi} s}{4}\right) .
\end{equation}\\
\begin{figure}[h]
  \centering    
\includegraphics[width=0.45\textwidth]{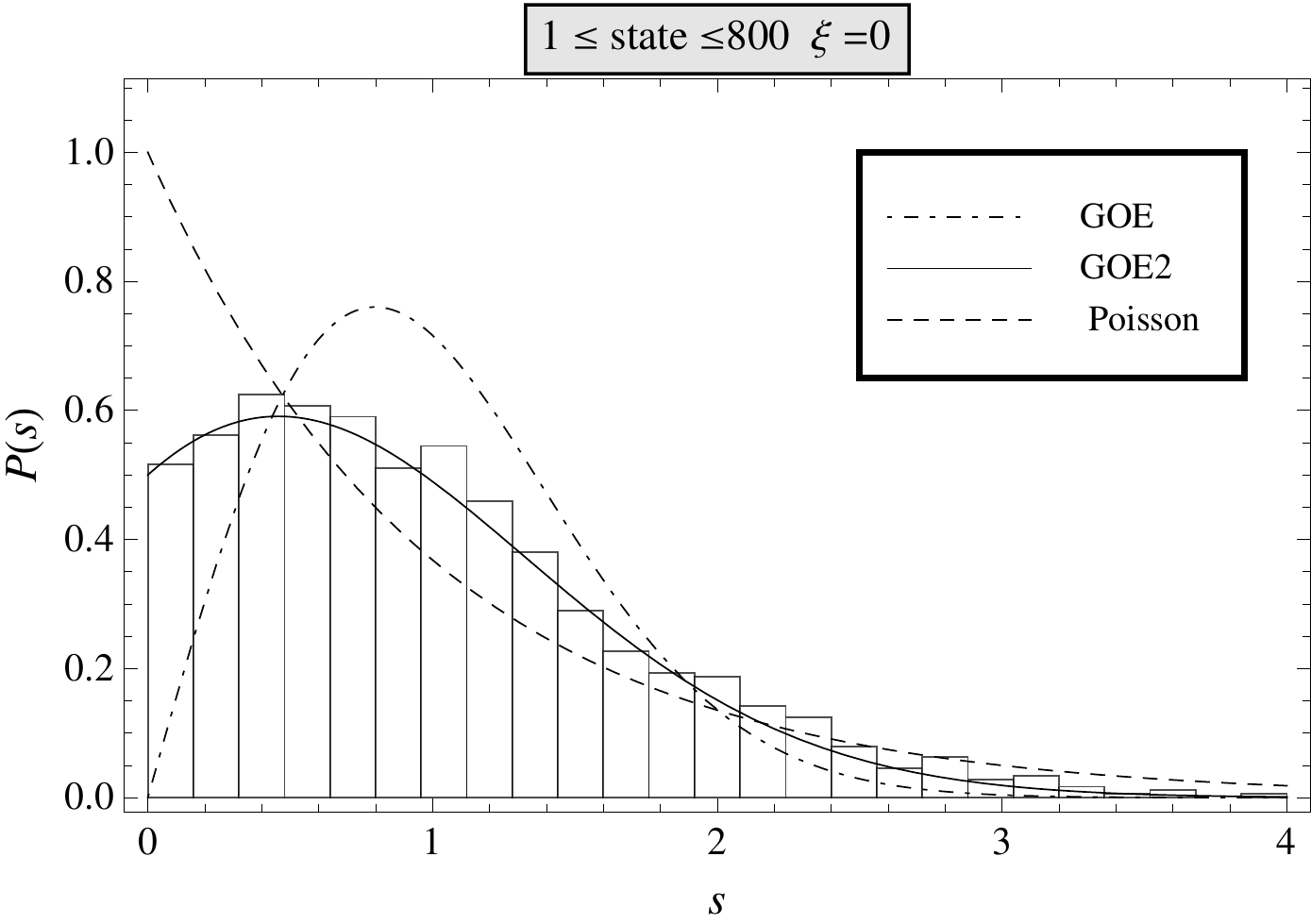}\hspace{0.1cm}
  \includegraphics[width=0.35\textwidth]{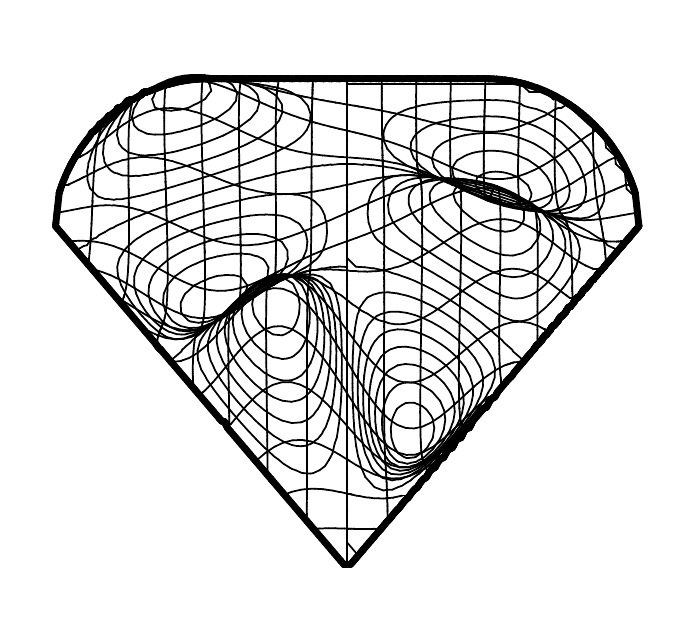}
  \caption[NNSD of the complete billiard.] {Nearest neighbor
    spacing distribution of the complete billiard. (left) The
    histogram of the level spacing distribution was built using the
    first 800 energy levels, the first one hundred states were not
    taken. The total number of energy levels computed using the finite
    difference method was $Q=11026$. (right) Ninth state of the complete diamond
    shaped billiard.}
  \label{goe2Fig}
\end{figure}

In Figure \ref{goe2Fig} it is shown the nearest neighbor spacing
distribution of the diamond shaped billiard which fits the $P_{2 GOE}$
(GOE2) distribution, as expected.
\begin{figure}[h]
  \centering
  \includegraphics[width=0.4\textwidth]{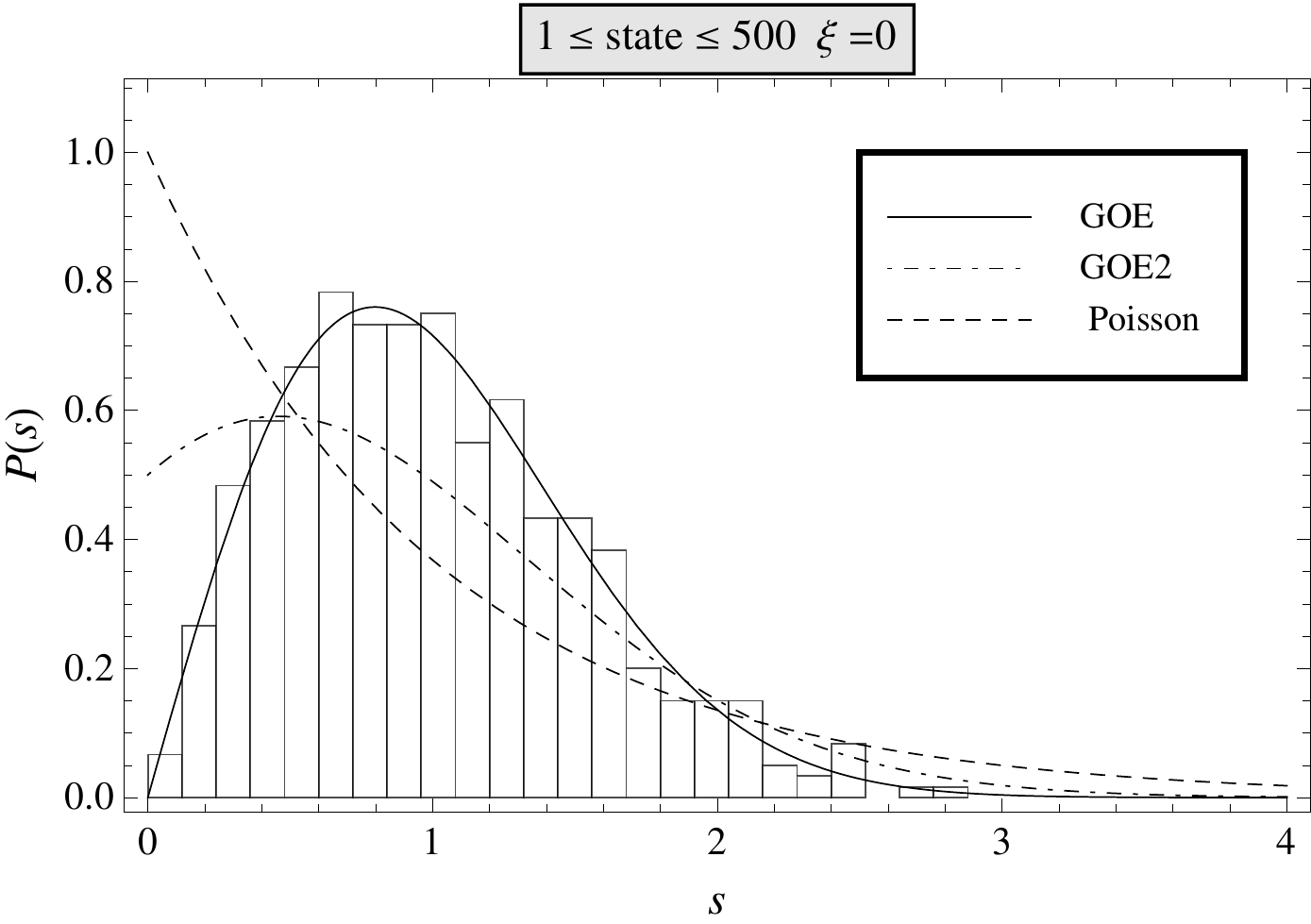}\hspace{0.1cm}
 \includegraphics[width=0.15\textwidth]{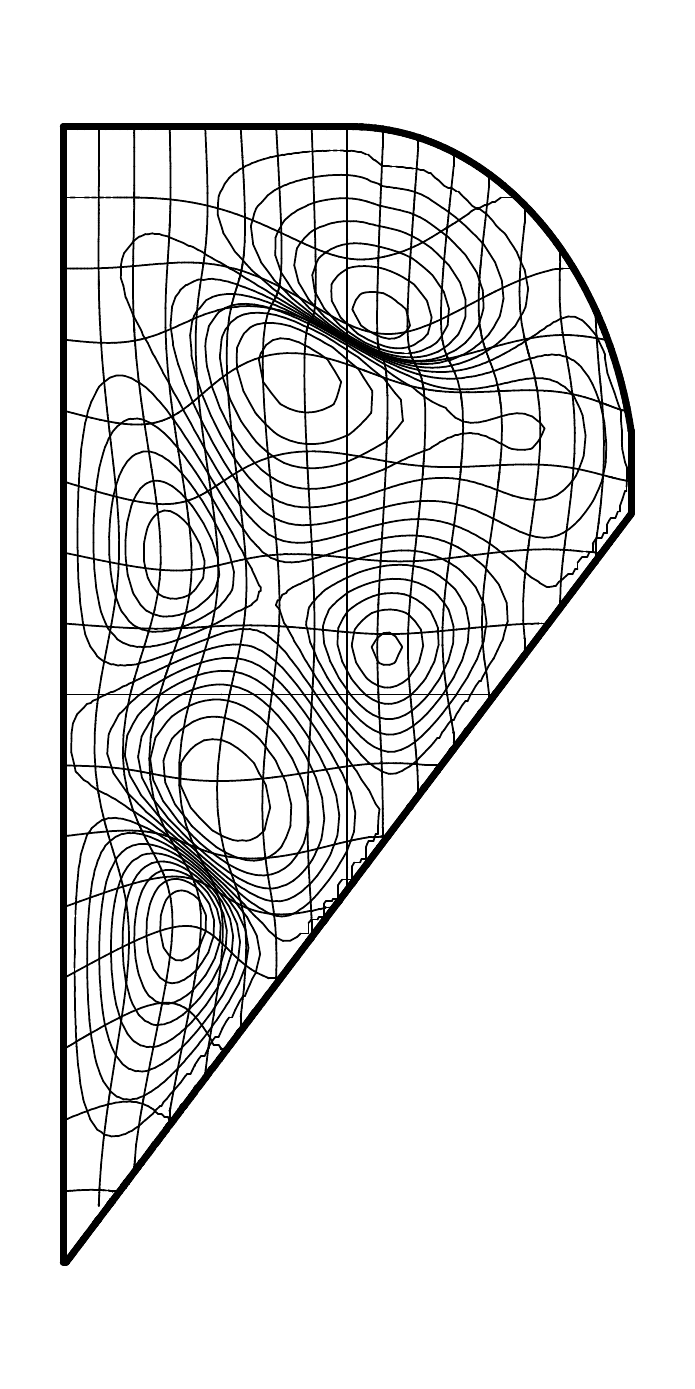}
  \caption[NNSD of the desymmetrized diamond shaped billiard.]%
          {Nearest neighbor spacing distribution of the
              desymmetrized diamond shaped billiard. (left) The
            histogram of the level spacing distribution was built
            using the first 500 energy levels. The total of energy
            levels computed using the finite difference method was
            $Q=11252$. (right) Ninth state of the desymmetrized diamond
            shaped billiard.}
  \label{goe1Fig}
\end{figure}

There are two ways to recover the GOE distribution: the first one is
to classify the energy levels according to the parity of the
eigenstates and the histogram is built with one of the two
sets. However, there is a disadvantage in this method, because it
requires to take approximately the half of the energy levels
computed. The second one consists in statistical study of the
corresponding desymmetrized billiard spectrum. In this case the
billiard is desymmetrized by taking a half of it for the mirror
symmetry of the diamond billiard. The level statistic effectively obeys a GOE distribution for the desymmetrized billiard. The result is
shown in Figure \ref{goe1Fig}.

\begin{figure}[h]
  \centering
  \includegraphics[scale=0.25]{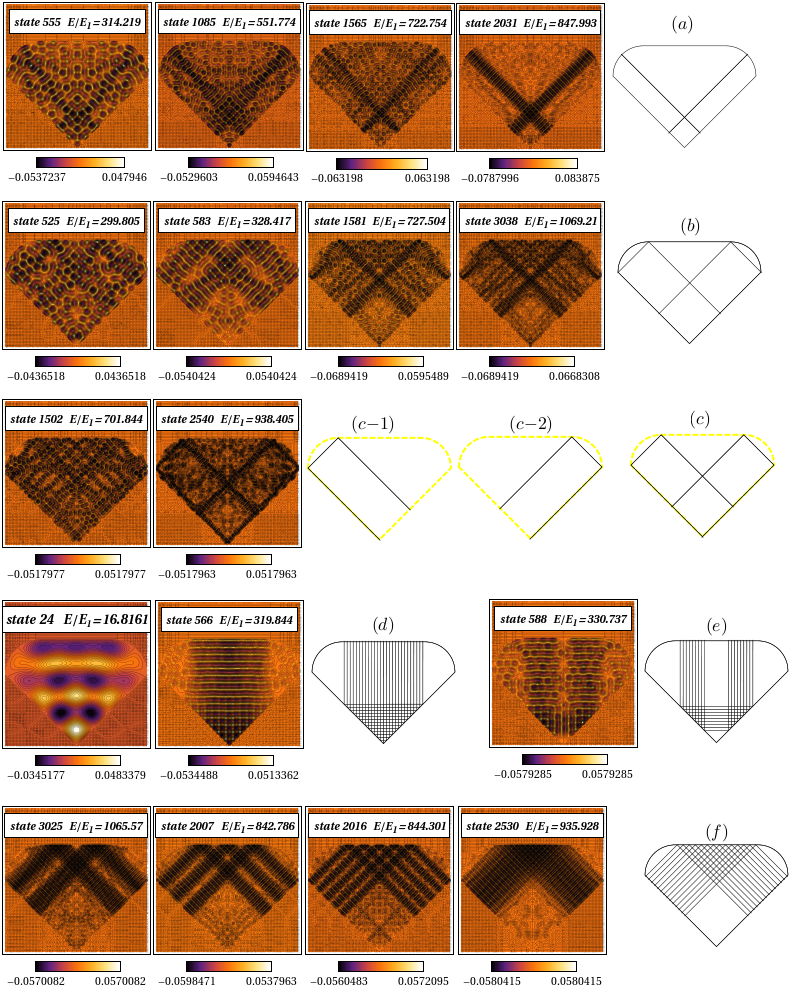} 
  \caption[Some scarred wavefunctions and bouncing ball states of the diamond shaped billiard.]%
  {Some scarred wavefunctions and bouncing ball states of the diamond shaped billiard.}
  \label{scarsAndBouncingFig}
\end{figure}
Another important feature of quantum chaotic billiards is the high
concentration of the wavefunction amplitude along the classical
periodic orbits. The phenomenon was initially observed by McDonald and
Kaufman \cite{McDonaldPaper}, and posteriorly in the Bunimovich
billiard by Heller \cite{HellerPaper}, who called it a
\textit{scar}. Using the analytic solution of the wavefunction it is
not possible to built a scar in the rectangular, circular or
equilateral triangle billiard. The scarring of wavefunction in
billiards does not appear in regular billiards, and it is exclusive
for the chaotic ones. As the quantum numbers are increased, we expect
to recover the classical characteristics of the system, which is, in
some sense, the idea behind the correspondence principle.  In a
chaotic billiard, the trajectories which evolve from an arbitrary
initial condition tend to fill the whole billiard, as consequence a
typical wavefunction should not have a significant
localization, which is the more common situation for the irregular
billiards. Nonetheless, for the special case of an unstable periodic
orbit, it is possible to find a high probability density underlying
such classical trajectory, as we may intuitively expect at least in
the semiclassical limit. Some scars and \textit{bouncing ball states}
are shown in Figure \ref{scarsAndBouncingFig}. The bouncing ball
states have a well defined momentum, but not position and we may
associate a set of classical periodic orbits to a single bouncing ball
state. In contrast, a scar is related to a single unstable periodic
orbit.\\
\begin{figure}[h]
  \centering
  \includegraphics[width=0.35\textwidth]{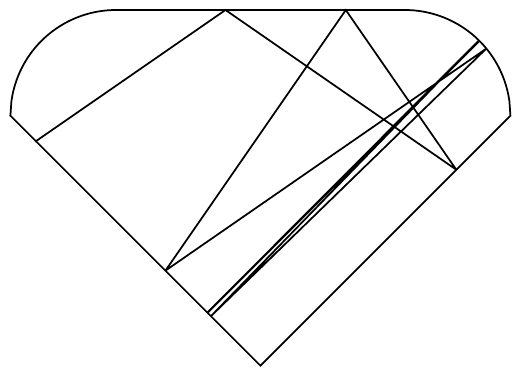}\hspace{0.5cm}
  \includegraphics[width=0.35\textwidth]{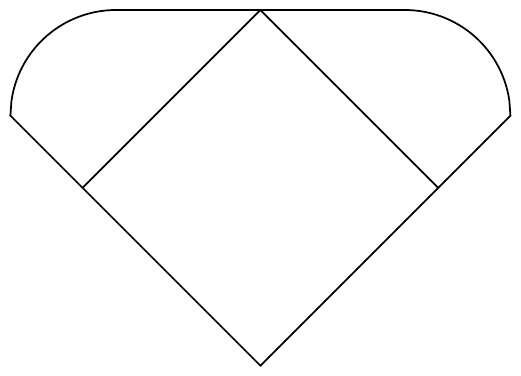}
  \caption[Orbits stability.]%
  {Orbits stability. (left) Unstable periodic orbit after 35 collisions, (right) stable periodic orbit after 1000 collisions.}
  \label{OrbitsStabilityFig}
\end{figure}\\ 
Example of scars are shown in Figure
\ref{scarsAndBouncingFig}-\textit{(a)} and
\ref{scarsAndBouncingFig}-\textit{(c)}. The stability of the orbits
with lowest period is shown in Figure \ref{OrbitsStabilityFig}.

\section{Quantum diamond billiard: time evolution of the state vector}
In this section the study will be limited to the time evolution of the
mean values. The time evolution of the state vector for quantum
Hamiltonian systems is well known $\mid\Psi\left(t\right)\rangle=e^{-i\frac{\hat{H}t}{\hbar}}\mid\Psi\left(0\right)\rangle .$ Expanding the state vector over the corresponding stationary states $\mid\Psi\left(\vec{r},0\right)\rangle = \sum_s c_s\mid\psi^{(s)}\rangle$, and evaluating on one arbitrary inner point of the lattice we find
\[
\left.\langle{\vec{r}}\mid\Psi\left(t\right)\rangle\right|_{\vec{r}=\vec{r}_u} = \Psi_u(t)= \sum_s c_s \exp{\left(-\frac{iE_st}{\hbar}\right)}\psi^{(s)}_u .
\]\\The components are given in the usual way, however we may use the lattice in order to compute them easily
\begin{equation}
c_s = \lim_{\delta^2\vec{r}\to 0}\sum_{u\in\mathfrak{D}}\psi^{(s)}_u\Psi_u(0)\delta^2\vec{r} .
\end{equation}
Since the eigenvectors of the Hamiltonian matrix, given by the equation (\ref{HamiltonianMatrix}), are real, then the complex conjugation has been dropped. If the number of inner points is large, we may use the last equation without the limit as a good approximation for the components computation. The same approach may be used for the computation of the several mean values involved in the uncertainty products for momentum and position. Using the index displacement transformations and central differences, the gradient evaluated at the point $u(i,j)$ may be expressed as 
\begin{equation}
\left(\vec{\nabla}\Psi\right)_u = \frac{1}{2}\left(\frac{\Psi_{u^+}(t)-\Psi_{u^-}(t)}{\delta x}, \frac{\Psi_{u_+}(t)-\Psi_{u_-}(t)}{\delta y}\right)\hspace{0.1cm},
\end{equation}
so the mean value for momentum takes the form
\begin{equation}
\langle\hat{\vec{p}}(t)\rangle = \frac{\hbar}{i} \lim_{\delta^2\vec{r}\to 0}\sum_{u\in\mathfrak{D}}\Psi_u(t)^*\left(\vec{\nabla}\Psi\right)_u\delta^2\vec{r} .
\end{equation}
Taking the real and imaginary part of the last equation we find
\[
\langle\hat{\vec{p}}(t)\rangle = \hbar \lim_{\delta^2\vec{r}\to 0}\sum_{u\in\mathfrak{D}}\left[\vec{\nabla},\Psi\right]_u(t)\delta^2\vec{r}
\]\\and
\begin{equation}
\lim_{\delta^2\vec{r}\to 0}\sum_{u\in\mathfrak{D}}\left\{\vec{\nabla},\Psi\right\}_u(t)\delta^2\vec{r} = 0
\label{averageMomentumEq}
\end{equation}
\\where the following expressions were defined
\begin{equation}
\begin{array}{l}
\displaystyle \left[\vec{\nabla},\Psi\right]_u(t):=\mathrm{Re}\left[\Psi_u(t)\right]\mathrm{Im}\left[\left(\vec{\nabla}\Psi\right)_u(t)\right]
				-\mathrm{Im}\left[\Psi_u(t)\right]\mathrm{Re}\left[\left(\vec{\nabla}\Psi\right)_u(t)\right] \mbox{ and } \vspace{0.5cm}\\
\displaystyle \left\{\vec{\nabla},\Psi\right\}_u(t):=\mathrm{Re}\left[\Psi_u(t)\right]\mathrm{Re}\left[\left(\vec{\nabla}\Psi\right)_u(t)\right]
				+\mathrm{Im}\left[\Psi_u(t)\right]\mathrm{Im}\left[\left(\vec{\nabla}\Psi\right)_u(t)\right] .		
\end{array}
\end{equation}
\\The condition in the equation (\ref{averageMomentumEq}) appears because the mean value is a real quantity, then for practical means, the imaginary part may be used to check the accuracy of the numerical integral evaluation. Similarly, for the mean value of the squared momentum we find

\begin{equation}
\langle\hat{\vec{p}}^{\hspace{0.05cm}2}(t)\rangle = -\hbar^2 \lim_{\delta^2\vec{r}\to 0}\sum_{u\in\mathfrak{D}}\left\{\vec{\nabla}^2,\Psi\right\}_u(t)\delta^2\vec{r}   	
\label{averageSquaredMomentumEq}
\end{equation}
with
\[
\lim_{\delta^2\vec{r}\to 0}\sum_{u\in\mathfrak{D}}\left[\vec{\nabla}^2,\Psi\right]_u(t)\delta^2\vec{r} = 0
\]
\\where the laplacian at the point $u(i,j)$ is
\begin{equation}
\left(\vec{\nabla}^2\Psi\right)_u = \frac{\Psi_{u^+}(t) + \Psi_{u^-}(t)}{\delta x^2} + \frac{\Psi_{u_+}(t) + \Psi_{u_-}(t)}{\delta y^2} - 2(\delta x^{-2}+\delta y^{-2})\Psi_u(t) .
\end{equation}
\\Finally, the average of an arbitrary function $f$ with only position dependence is
\begin{equation}
\langle f(\hat{\vec{r}})\rangle(t) =  \lim_{\delta^2\vec{r}\to 0}\sum_{u\in\mathfrak{D}}\mid\Psi_u(t)\mid^2 f_u .
\end{equation}

\begin{figure}[h]
  \centering
  \includegraphics[width=0.3\textwidth]{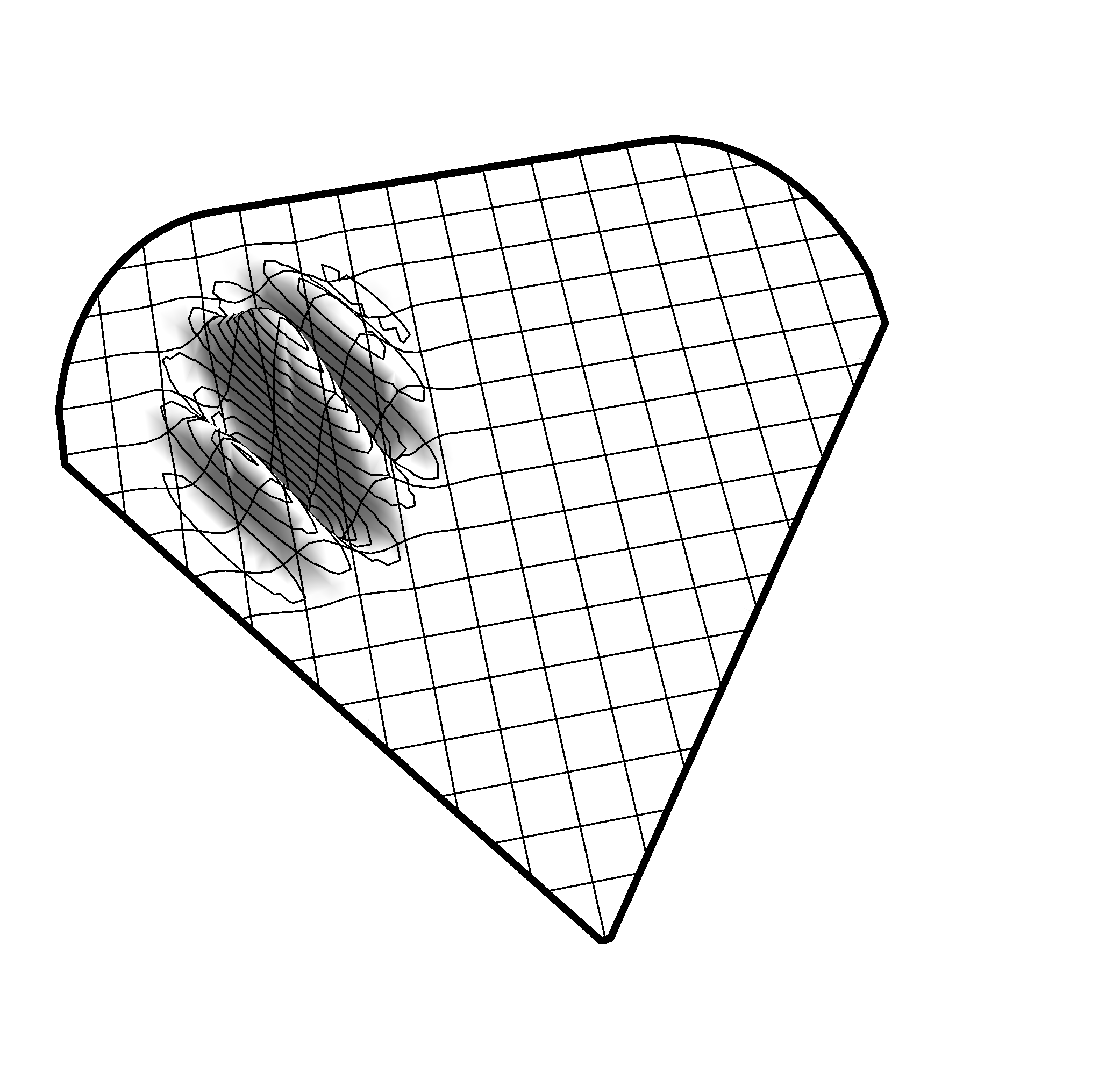}\hspace{0.1cm} 
  \includegraphics[width=0.35\textwidth]{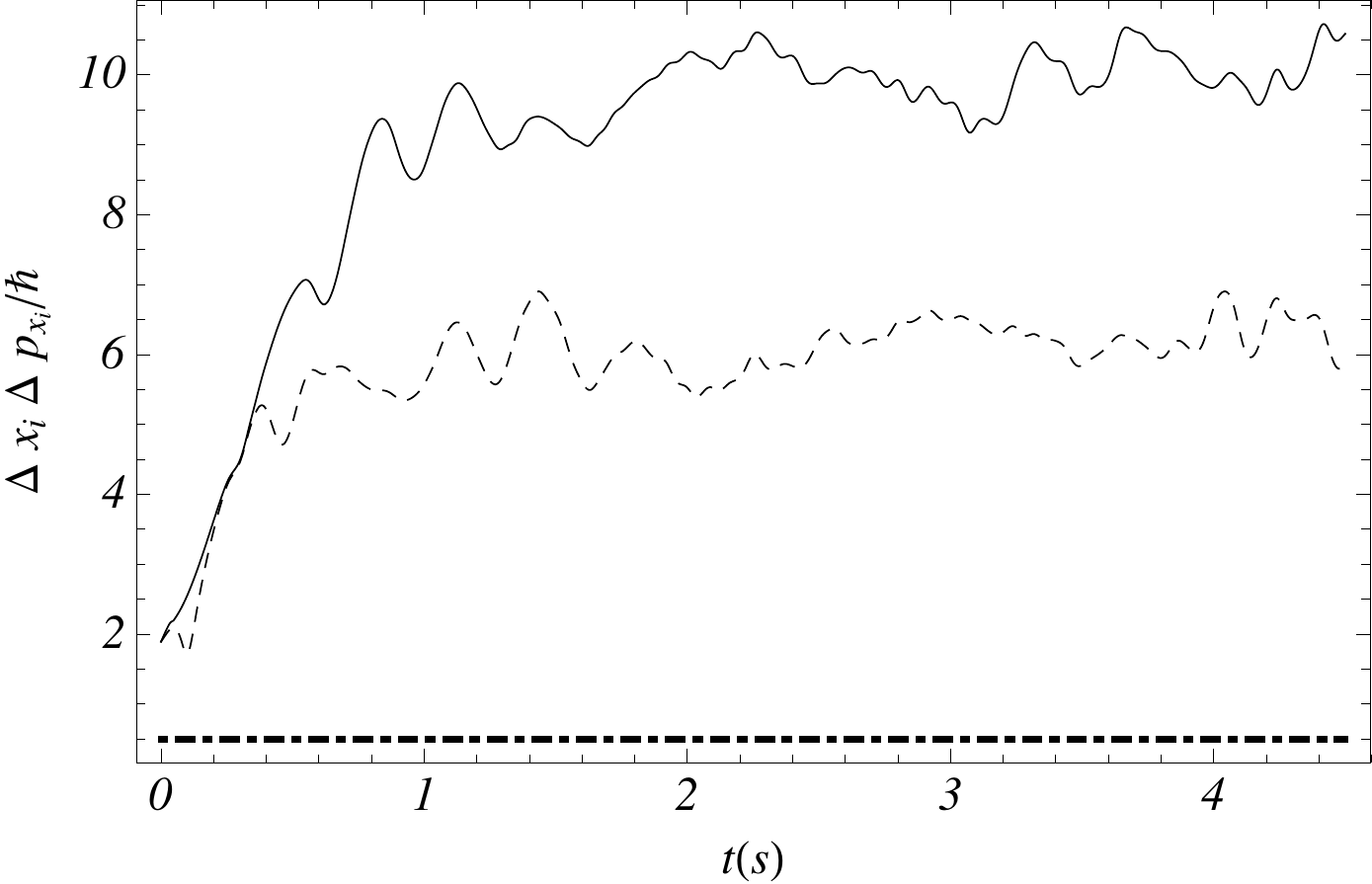}\\ 
  \includegraphics[width=0.5\textwidth]{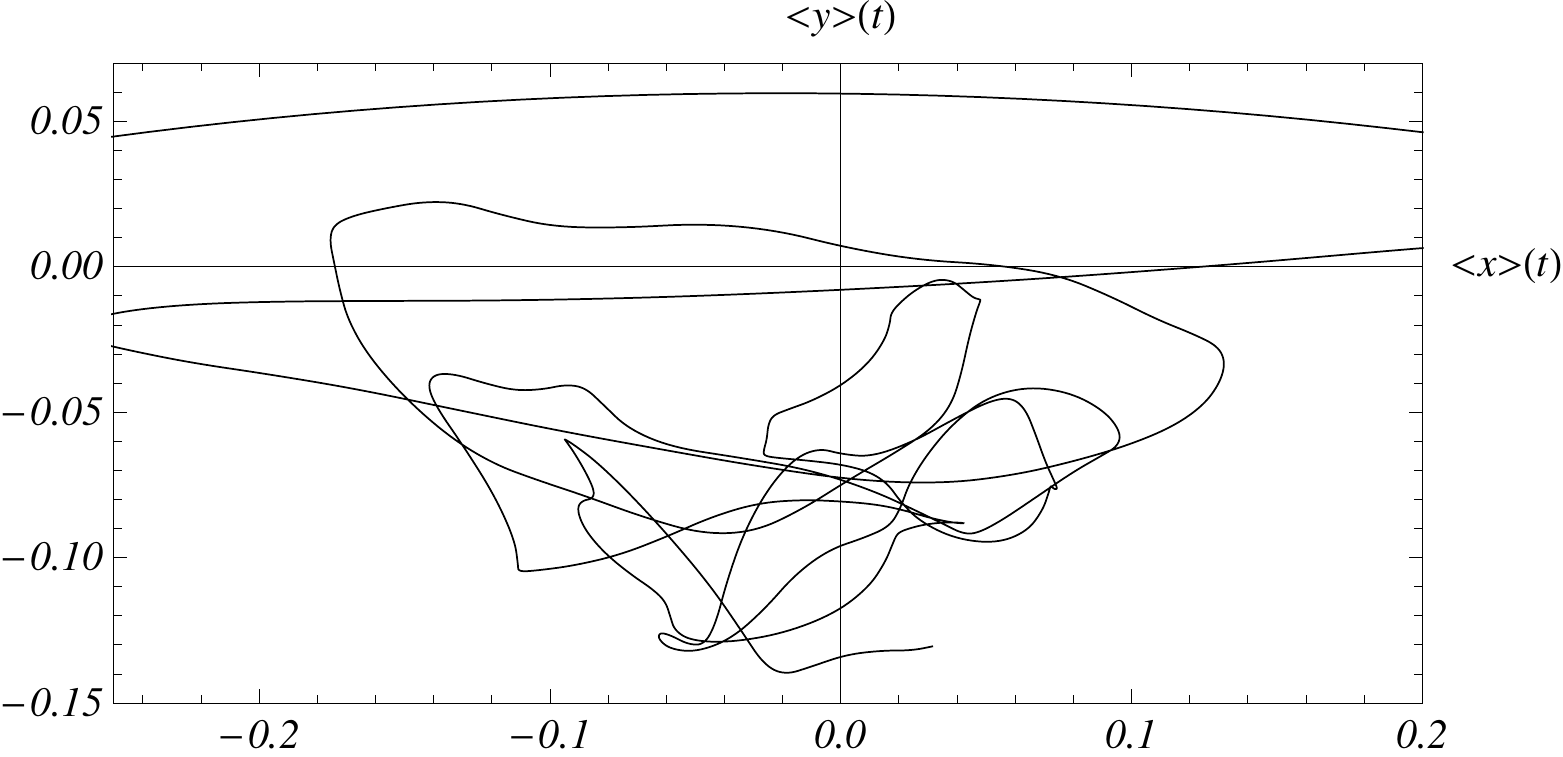}      
  \caption[Gaussian wave packet evolution.]
  {Gaussian wave packet evolution. (Top). (left) Initial state. (right)
    Uncertainty products: $\Delta x \Delta p_x$ (solid line) and
    $\Delta y \Delta p_y$ (dashed line). The dot dashed line is the
    minimum uncertainty value $\Delta x\Delta p_{x}=\hbar/2$
    (dot-dashed line). (Bottom). Trajectory of the position operator
    expectation values.}
  \label{uncertaintyFig}
\end{figure}

\begin{figure}[h]
  \centering
  \includegraphics[width=0.15\textwidth]{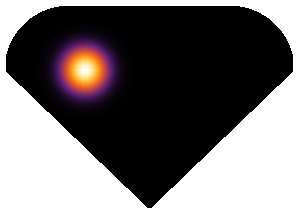}
  \includegraphics[width=0.15\textwidth]{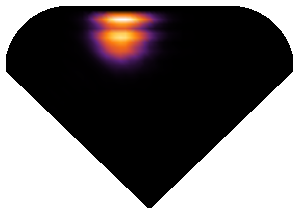}       
  \includegraphics[width=0.15\textwidth]{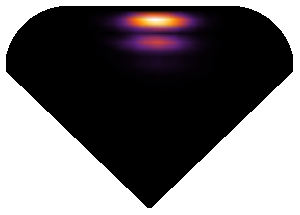}       
  \includegraphics[width=0.15\textwidth]{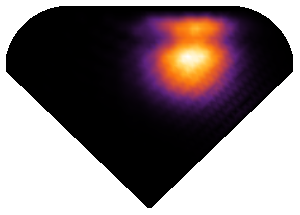}       
  \includegraphics[width=0.15\textwidth]{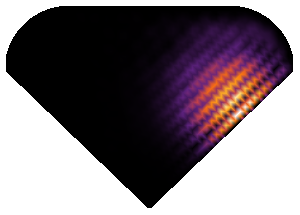}\\       
  \includegraphics[width=0.15\textwidth]{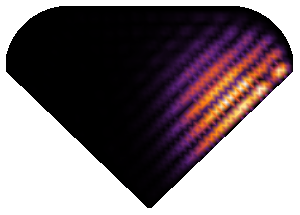}       
  \includegraphics[width=0.15\textwidth]{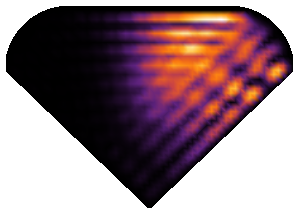}       
  \includegraphics[width=0.15\textwidth]{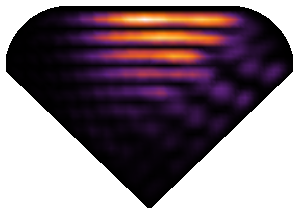}       
  \includegraphics[width=0.15\textwidth]{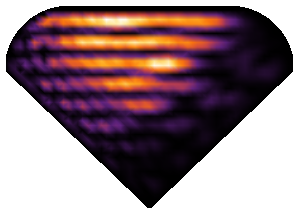}       
  \includegraphics[width=0.15\textwidth]{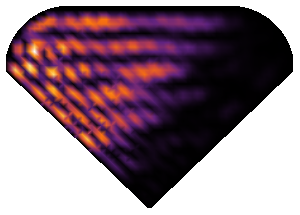}\\       
  \includegraphics[width=0.15\textwidth]{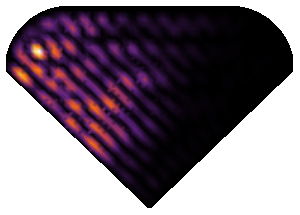}       
  \includegraphics[width=0.15\textwidth]{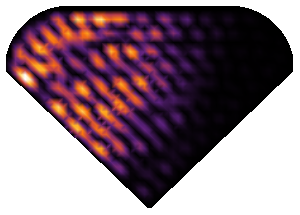}       
  \includegraphics[width=0.15\textwidth]{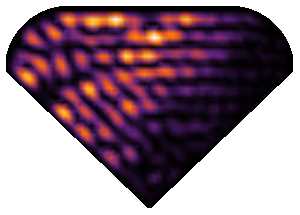}       
  \includegraphics[width=0.15\textwidth]{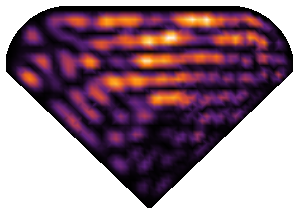}       
  \includegraphics[width=0.15\textwidth]{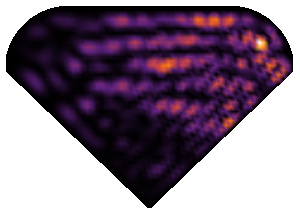}\\ 
  \includegraphics[width=0.15\textwidth]{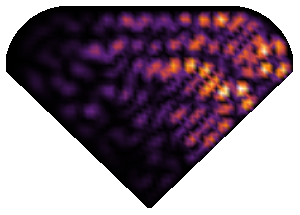}       
  \includegraphics[width=0.15\textwidth]{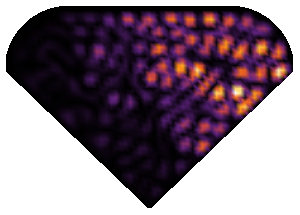}       
  \includegraphics[width=0.15\textwidth]{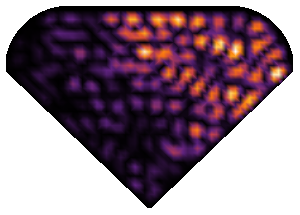}       
  \includegraphics[width=0.15\textwidth]{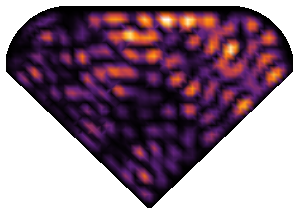}       
  \includegraphics[width=0.15\textwidth]{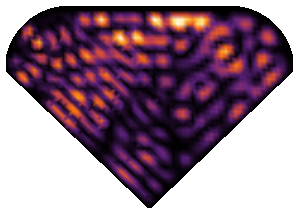}  
  \caption[Probability density distribution of the billiard prepared in a Gaussian wave packet.]%
  {Probability density distribution of the billiard prepared in a Gaussian wave packet. The packet was initially placed in $\vec{r}_o=(-R,0)$. The wave vector was pointed to $\pi/4$ rad respect the $x$-axis. Classically, the particle should describe the 4-periodic orbit shown in Figure \ref{OrbitsStabilityFig}; nonetheless, after the second collision the corresponding state is highly delocalized. The particle tends to remain on a bouncing ball state after the second collision (see Figure \ref{scarsAndBouncingFig}-(f)); however, the next collisions increase the delocalization and the evolution turns non-coherent.}
  \label{wavepacketEvolution}
\end{figure}
The time evolution of the position mean values of the system
prepared in a well localized initial state at $\vec{r}_o = (x_o,y_o)$
using a Gaussian wave packet
\begin{equation}
  \langle\vec{r}\mid \Psi (0) \rangle = \frac{e^{-\frac{(x-x_o)^2}{4\sigma_x^2}-\frac{(y-y_o)^2}{4\sigma_y^2} + i\left(\kappa_x x+\kappa_y y\right)}}{2\pi\sigma_x\sigma_y}  
\end{equation}
is shown in Figure
\ref{uncertaintyFig}. $\vec{\kappa}=(\kappa_x,\kappa_y)$ is the
wave vector, $\sigma_x$ and $\sigma_y$ are the standard deviation on
along $x$ or $y$ respectively. As seen in
Figure~\ref{wavepacketEvolution}, the wavepacket is destroyed after a
few collisions. However, this is not a consequence of the classical chaotic behaviour and such irregular dynamics may be
attributed to a non-coherent preparation of the initial state. This is
checked in the evolution of the uncertainty products. The system must
be prepared in a coherent state in order to reduce the uncertainty
products to their minimum value $\hbar/2$; nevertheless, we do not
have a general analytical expression for the coherent states of
billiards, even in simple cases such as a particle in a rectangular
box.\\
Classical chaos of an specific system often emerges from its nonlinear
nature. Nevertheless, the classical and quantum billiard systems are
an exception of this rule because of the absence of nonlinear terms in
their governing equations. Indeed, the difficulty for quantum chaos
determination does not lie in this lack of nonlinearity rather the
problem resides in the difficulty to find a correspondence between the
classical and quantum behaviour far from the classical limit when the
classical system evolves chaotically. The question is not solved by
simply proving the Bohigas-Giannoni-Schmit conjecture because
the nearest neighbor spacing distribution is just a semiclassical
result. The analysis of the position operator expectation value is an
alternative to study the correspondence between the classical and
quantum system in an irregular regime. However, this approach
frequently is not successful because the quantum system evolves in a
non-coherent way when its classical counterpart is chaotic as we show
in this writing. This would be reason, for which the quantum
Hamiltonian system sensibility has been sometimes studied by
perturbing the Hamiltonian instead of by changing the initial state
\cite{PerturbedStadium}.

\section{Concluding remarks}

The classical and quantum diamond billiard was studied through some
quantities. In particular, we calculate the entropy, the Lyapunov
exponent and some trajectories for the classical problem. The classical chaotic
behaviour emerges fast with small modifications on the boundary of the
regular equilateral billiard ($\xi=1$). The entropy and the Lyapunov
exponent grow when a half of stadium is introduced in one side of the
triangular billiard. If the control parameter is set far enough from
one, say in the interval $0.8 \leq \xi \leq 1$, then the entropy
practically is a $70\%$ of $S_{max}$. This percentage is relative far
from its maximum and it occurs because the diamond billiard does not
have dispersive frontiers as other billiards e.g. Sinai
billiard. Nonetheless, it is enough to ensure a great irregularity in
the classical trajectories when the entropy is about
$0.7S_{max}$.

The finite difference method provides a way
to solve the quantum problem. The diamond shape billiard has a mirror
reflecting symmetry. Because of this, the energy levels split in two
different symmetry classes according to the wavefunction parity. As
consequence, $P(s)$ for the complete billiard is given by a GOE2
distribution. If diamond billiard is desymmetrized, then
the level statistics follows a GOE distribution. On the other hand, the
classical behaviour is regular when the control parameter is set to
one and the distribution is Poissonian. Therefore, the results are
according to the Bohigas-Giannoni-Schmit conjecture. We found scarred
states in the quantum diamond billiard, as well as bouncing ball
states with their corresponding set of classical stable periodic
orbits. These results are in agreement with previous work in the field
for other Hamiltonian systems. In the last section, a practical way to
compute the time evolution of the state vector was described and the
lattice previously built in the finite difference method
implementation was used for this aim.

\section*{acknowledgments}
This work was supported by Facultad de Ciencias de la Universidad de
los Andes, and ECOS NORD/COLCIENCIAS-MEN-ICETEX. D. L. Gonz\'alez was
supported by the NSF-MRSEC at the University of Maryland, Grant
No. DMR 05-20471, and a DOE-BES-CMCSN grant, with ancillary support
from the Center for Nanophysics and Advanced Materials (CNAM).

\end{document}